\documentclass{vldb}
\usepackage{graphicx}
\usepackage{balance}  
\usepackage{epsfig}
\usepackage{subfigure}
\usepackage{amsmath}
\usepackage{float}
\usepackage{amssymb}
\usepackage{latexsym}
\usepackage{epic}
\usepackage{url}
\usepackage{algorithm}
\usepackage[noend]{algorithmic}
\usepackage{tabularx}
\usepackage{times}
\usepackage{color}

\newcommand{\eat}[1]{}

\newtheorem{theorem}{Theorem}
\newtheorem{lemma}{Lemma}

\newtheorem{definition}{Definition}

\long\def\comment#1{}

\begin{document}


\title{Low-Rank Mechanism: Optimizing Batch Queries \\under Differential Privacy}



%
%
%
%

\numberofauthors{1}

%
\author{
Ganzhao Yuan$^1$~~Zhenjie Zhang$^2$~~Marianne Winslett$^{2,3}$~~Xiaokui Xiao$^4$~~Yin Yang$^{2}$~~Zhifeng Hao$^1$ \\
\and $^1$School of Computer Science \& Engineering \\
South China University of Technology\\
\{y.ganzhao,mazfhao\}@scut.edu.cn \\
\and $^2$Advanced Digital Sciences Center \\
Illinois at Singapore Pte. Ltd.\\
\{zhenjie,yin.yang\}@adsc.com.sg \\
\and $^3$Dept. of Computer Science \\
University of Illinois at Urbana-Champaign\\
winslett@illinois.edu\\
\and $^4$School of Computer Engineering \\
Nanyang Technological University \\
xkxiao@ntu.edu.sg
}
\maketitle


\begin{abstract}
Differential privacy is a promising privacy-preserving paradigm for
statistical query processing over sensitive data. It works by
injecting random noise into each query result, such that it is
provably hard for the adversary to infer the presence or absence of
any individual record from the published noisy results. The main
objective in differentially private query processing is to maximize
the accuracy of the query results, while satisfying the privacy
guarantees. {Previous work, notably \cite{LHR+10}, has suggested that with an appropriate strategy, processing a batch of correlated
queries as a whole achieves considerably higher accuracy
than answering them individually. However, to our knowledge there is currently no practical solution to find such a strategy for an arbitrary query batch; existing methods either return strategies of poor quality (often worse than naive methods) or require prohibitively expensive computations for even moderately large domains.
}
Motivated by this, we propose the \emph{Low-Rank Mechanism} (LRM), the first practical
differentially private technique for answering batch queries with
high accuracy, based on a \emph{low rank approximation} of the
workload matrix. We prove that the accuracy provided by LRM is close
to the theoretical lower bound for any mechanism to answer a batch
of queries under differential privacy. Extensive experiments using
real data demonstrate that LRM consistently outperforms
state-of-the-art query processing solutions under differential
privacy, by large margins.
\end{abstract}

\section{Introduction}\label{sec:intro}

Differential privacy \cite{DMNS06} is an emerging paradigm for publishing statistical information over sensitive data, with strong and rigorous guarantees on individuals' privacy. Since its proposal, differential privacy has attracted extensive research efforts, such as cryptography \cite{DMNS06}, algorithms \cite{DRV10,HT10,MT07}, databases \cite{DWHL11,HRMS10,LHR+10,RN10,XBHG11,XWG10,XZXYY11}, data mining \cite{BLS+10,FS10} and machine learning \cite{BLR08,CMS11,RBHT09}. The main idea of differential privacy is to inject random noise into aggregate query results, such that the adversary cannot infer, with high confidence, the presence or absence of any given record $r$ in the dataset, even if the adversary knows all other records in the dataset except for $r$. This paper follows a popular definition of differential privacy, called $\epsilon$-differential privacy, in which the adversary's maximum confidence in inferring private information is controlled by a user-specified parameter $\epsilon$ called the \emph{privacy budget}. Given $\epsilon$, the main goal of query processing under $\epsilon$-differential privacy is to maximize the utility/accuracy of the (noisy) query answers, while satisfying the above privacy requirements.

\begin{figure} [htb]
\centering
\begin{tabular}{cc}
{\includegraphics[height=0.15\textwidth]{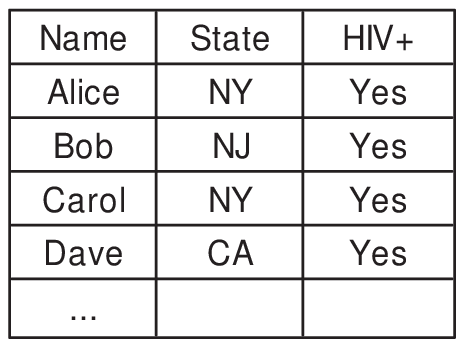}} &
{\includegraphics[height=0.15\textwidth]{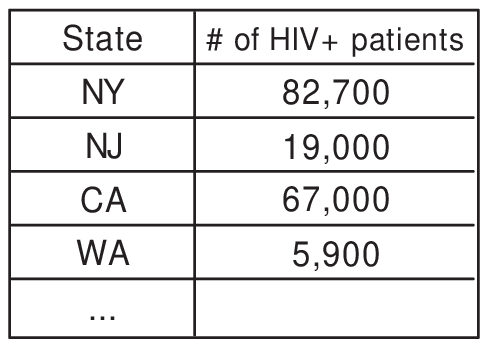}} \\
(a) Patient records & (b) Statistics on HIV+ patients
\end{tabular}
\caption{Example medical record database}
\label{fig:intro-example}
\end{figure}

This work focuses on a common class of queries called \emph{linear counting queries}, which is the basic operation in many statistical analyses. Similar ideas apply to other types of linear queries, e.g., linear sums. Figure \ref{fig:intro-example}(a) illustrates an example electronic medical record database, where each record corresponds to an individual. Figure \ref{fig:intro-example}(b) shows the exact number of HIV+ patients in each state, which we refer to as \emph{unit counts}. A linear counting query in this example can be any linear combination of the unit counts. For instance, let $x_{NY}$, $x_{NJ}$, $x_{CA}$, $x_{WA}$ be the patient counts in states NY, NJ, CA, and WA respectively; one possible linear counting query is $x_{NY}+x_{NJ}+x_{CA}+x_{WA}$, which computes the total number of HIV+ patients in the four states listed in our example. Another example linear counting query is $x_{NY}/19+x_{NJ}/8+x_{CA}/37$, which calculates the weighted average of patient counts in states NY, NJ and CA, with weights set according to their respective population sizes. In general, we are given a database with $n$ unit counts, and a batch $QS$ of $m$ linear counting queries. The goal is to answer all queries in $QS$ under $\epsilon$-differential privacy, and maximize the expected overall accuracy of the queries.

Straightforward approaches to answering a batch of linear counting queries usually lead to sub-optimal result accuracy. One na\"{\i}ve solution, referred to as \emph{noise on queries} (\emph{NOQ}), is to process each query independently, e.g., using the Laplace Mechanism \cite{DMNS06}. This method fails to exploit the \emph{correlations} between different queries. Consider a batch of three different queries $q_1=x_{NY}+x_{NJ}+x_{CA}+x_{WA}$, $q_2=x_{NY}+x_{NJ}$, $q_3=x_{CA}+x_{WA}$. Clearly, the three queries are correlated since $q_1=q_2+q_3$. Thus, an alternative strategy for answering these queries is to process only $q_2$ and $q_3$, and use their sum to answer $q_1$. As will be explained in Section 3, the amount of noise added to query results depends upon the \emph{sensitivity} of the query set, which is defined as the maximum possible total change in query results caused by adding or removing a single record in the original database. In our example, the sensitivity of the query set $\{q_2, q_3\}$ is 1, because adding/removing a patient record in Figure \ref{fig:intro-example}a affects at most one of $q_2$ and $q_3$ (i.e., $q_2$ if the record is associated with state NY or NJ, and $q_3$ if the state is CA or WA), by exactly 1. On the other hand, the query set $\{q_1, q_2, q_3\}$ has a sensitivity of $2$, since a record in the above 4 states affects both $q_1$ and one of $q_2$ and $q_3$. According to the Laplace mechanism, the variance of the added noise to each query is $2\Delta^2/\epsilon^2$, where $\Delta$ is the sensitivity of the query set, and $\epsilon$ is the user-specified privacy budget. Therefore, processing $\{q_1, q_2, q_3\}$ directly incurs a noise variance of $8/\epsilon^2$ for each query; on the other hand, executing $\{q_2, q_3\}$ leads to noise variance of $2/\epsilon^2$ for each of $q_2$ and $q_3$, and their sum $q_1=q_2+q_3$ has a noise variance of $2\times2/\epsilon^2=4/\epsilon^2$. Clearly, the latter method obtains higher accuracy for all queries.

Another simple solution, referred to as \emph{noise on data} (\emph{NOD}), is to process each unit count under differential privacy, and combine them to answer the given linear counting queries. Continuing the example, this method computes the noisy counts for $x_{NY}$, $x_{NJ}$, $x_{CA}$ and $x_{WA}$, and uses their linear combinations to answer $q_1$, $q_2$, and $q_3$. This approach overlooks the correlations between different unit counts. In our example, $x_{NY}$ and $x_{NJ}$ (and similarly, $x_{CA}$ and $x_{WA}$) are either both present or both absent in every query, and, thus, can be seen as a single entity. Processing them as independent queries incurs unnecessary accuracy costs when re-combining them. In the example, NOD adds noise with variance $2/\epsilon^2$ to each unit count, and their combinations to answer $q_1$, $q_2$, and $q_3$ have noise variance $8/\epsilon^2$, $4/\epsilon^2$ and $4/\epsilon^2$ respectively. NOD's result utility is also worse than the above-mentioned strategy of processing $q_2$ and $q_3$, and adding their results to answer $q_1$.

In general, the query set $QS$ may exhibit complex correlations among different queries and among different unit counts. As a consequence, it is non-trivial to obtain the best strategy to answer $QS$ under differential privacy. For instance, consider the following query set:
\begin{eqnarray*}
q_1&=&2x_{NJ}+x_{CA}+x_{WA}\\
q_2&=&x_{NJ}+2x_{WA}\\
q_3&=&x_{NY}+2x_{CA}+2x_{WA}
\end{eqnarray*}

NOQ is clearly a poor choice, since it incurs a sensitivity of 5 (e.g., a record of state WA affects $q_1$ by 1, and $q_2$ and $q_3$ by 2 each). The sensitivity of NOD remains 1, and it answers $q_1$, $q_2$, and $q_3$ with noise variance $12/\epsilon^2$, $10/\epsilon^2$ and $18/\epsilon^2$ respectively, leading to a sum-square error (SSE) of $40/\epsilon^2$. The optimal strategy in terms of SSE in this case computes the noisy results of $x_{NJ}$ and $x_{WA}$, as well as $q'_1=x_{NY}/3+x_{CA}$, and $q'_2=2x_{NY}/3$. Then, it obtains the results for $q_1$, $q_2$, and $q_3$ as follows.
\begin{eqnarray*}
q_1&=&q'_1+2x_{NJ}+x_{WA}-q'_2/2\\
q_2&=&x_{NJ}+2x_{WA}\\
q_3&=&2q'_1+2x_{WA}+q'_2/2
\end{eqnarray*}

The sensitivity of the above method is also 1, and it answers $q_1$, $q_2$, and $q_3$ with noise variance $12.5/\epsilon^2$, $10/\epsilon^2$ and 16.5/$\epsilon^2$ respectively, resulting an SSE of 39/$\epsilon^2$. Observe that the there is no simple pattern in the query set or the optimal strategy. Since there is an infinite space of possible strategies, searching for the best one is a challenging problem.


{Li et al. \cite{LHR+10} first formalize the above observations (i.e., answering a correlated query set with an effective strategy) into the \emph{matrix mechanism}. However, applying the matrix mechanism in practice remains hard, because there is currently no effective solution to find a good strategy for an arbitrary query set. The only known strategy-searching methods described in \cite{LHR+10} are either inefficient (which incur prohibitively high computational costs for even moderately large domains), or ineffective (which rarely obtain strategies that outperform naive methods NOD/NOQ). Motivated by this, we
propose the first practical realization of the matrix mechanism, called the \emph{low-rank mechanism} (\emph{LRM}), based on the theory of low-rank matrix approximation.}
We prove that the accuracy provided by LRM is within a constant
factor of the theoretical lower bound established in \cite{HT10}.
Extensive experiments demonstrate that LRM significantly outperforms
existing solutions in terms of result accuracy, sometimes by orders
of magnitude.

The rest of the paper is organized as follows. Section
\ref{sec:related} reviews previous studies on differential privacy.
Section \ref{sec:pre} provides formal definitions for our problem.
Section \ref{sec:formulation} presents the mechanism formulation of
LRM, and analyzes its optimality. Section \ref{sec:opt} discusses
how to solve the optimization problem in LRM. Section \ref{sec:exp}
verifies the superiority of our proposal through an extensive
experimental study. Finally, Section \ref{sec:concl} concludes the
paper.

\section{Related Work}\label{sec:related}

Section \ref{sec:related:general} surveys general purpose mechanisms for enforcing differential privacy. Section \ref{sec:related:matrix} presents our main competitor, the matrix mechanism \cite{LHR+10}.

\subsection{Differential Privacy Mechanisms}\label{sec:related:general}

Differential privacy was first formally presented in \cite{DMNS06},
though some previous studies have informally used similar models,
e.g., \cite{DN03}. The Laplace mechanism \cite{DMNS06} is the first
generic mechanism for enforcing differential privacy, which works
when the output domain is a multi-dimensional Euclidean space.
McSherry and Talwar \cite{MT07} propose the exponential mechanism, which applies to
any problem with a measurable output space. The generality of
the exponential mechanism makes it an important tool in the design of many other
differentially private algorithms, e.g., \cite{CPS+12,XZXYY11,MT07}.

Linear query processing is of particular interest in both the theory and
database communities, due to its wide range of applications. To minimize the error of linear queries under differential privacy requirements, several
methods try to build a synopsis of the original database, such as
Fourier transformations \cite{RN10}, wavelets \cite{XWG10} and hierarchical trees \cite{HRMS10}. By publishing a noisy synopsis under $\epsilon$-differential privacy, these methods
are capable of answering an arbitrary number of linear queries.
However, most of these methods obtain good accuracy only when the
query selection criterion is a continuous range; meanwhile, since
these methods are not workload-aware, their performance for a
specific workload tends to be sub-optimal.

The compressive mechanism \cite{LZMY11} reduces the amount of noise necessary to satisfy differential privacy, by utilizing the sparsity of the dataset under certain transformations. The main idea is to use a technique called compressive sensing to compress a sparse representation of the data into a compact synopsis, and inject noise into the much smaller synopsis instead of the original data. After that, the method reconstructs the original data by applying the decoding algorithm of compressive sensing to the noisy synopsis. The result provides significantly higher utility, while satisfying differential privacy requirements.

Several theoretical studies have derived lower bounds for the noise level for processing linear queries under differential privacy. Notably, Dinur and Nissim \cite{DN03} prove that any perturbation mechanism with maximal noise of scale $\mathcal{O}(n)$ cannot possibly preserve personal privacy, if the adversary is allowed to ask all possible linear queries, and has exponential computation capacity. By reducing the computation capacity of the adversary to polynomial-bounded Turing machines, they show that an error scale $\Omega(\sqrt{n})$ is necessary to protect any individual' privacy.

More recently, Hardt and Talwar \cite{HT10} have significantly
tightened the error lower bound for answering a batch of linear
queries under differential privacy. Given a batch of $m$ linear
queries, they prove that any $\epsilon$-differential privacy
mechanism leads to squared error of at least
$\Omega(\epsilon^{-2}m^3Vol(W))$, where $Vol(W)$ is the volume of
the convex body obtained by transforming the $\mathcal{L}_1$-unit
ball into $m$-dimensional space using the linear transformations in
the workload $W$. They also propose a mechanism for differential
privacy whose error level almost reaches this lower bound. However,
their mechanism relies on uniform sampling in a high-dimensional
convex body, which, although it theoretically takes polynomial time,
is too expensive to be of practical use. This paper extends their
analysis to low-rank workload matrices.

Besides linear queries, differential privacy is also applicable to more complex queries in various research areas, due to its strong privacy guarantee.
In the field of data mining, Friedman and Schuster \cite{FS10} propose the first algorithm for building a decision tree under differential privacy. Mohammed et al. \cite{MCF+11} study the same problem, and propose an improved solution based on a generalization strategy coupled with the exponential mechanism. Ding et al. \cite{DWHL11} investigate the problem of differentially private data cube publication. They present a randomized materialized view selection algorithm, which reduces the overall error, and preserves data consistency.

In the database literature, a plethora of methods have been proposed to optimize the accuracy of differentially private query processing. Cormode et al. \cite{CPS+12} investigate the problem of multi-dimensional indexing under differential privacy, with the novel idea of assigning different amounts of privacy budget to different levels of the index. Xu et al. \cite{XZXYY11} optimize the procedure of building a differentially private histogram, with an interesting combination of a dynamic programming algorithm for optimal histogram computation and the exponential mechanism.

Differential privacy is also becoming a hot topic in the machine learning community, especially for learning tasks involving sensitive information, e.g., medical records. In \cite{CMS11}, Chaudhuri et al. propose a generic differentially private learning algorithm, which requires strong convexity of the objective function. Rubinstein et al. \cite{RBHT09} study the problem of SVM learning on sensitive data, and propose an algorithm to perturb the kernel matrix with performance guarantees, when the loss function satisfies the $l$-Lipschitz continuity property. General differential privacy techniques have also been applied to real systems, such as network trace analysis \cite{MM10} and private recommender systems \cite{MM09}.

\subsection{Matrix Mechanism}\label{sec:related:matrix}

{Li et al. \cite{LHR+10} propose the matrix mechanism, which formalizes the intuition that a batch of correlated linear queries can be answered more accurately under differential privacy, by processing a different set of queries (called the \emph{strategy}) and combining their results. Specifically, given a workload of linear
queries, the matrix mechanism first constructs a \emph{workload
matrix} $W$ of size $m$$\times$$n$, where $m$ is the number of
queries, and $n$ is the number of unit counts. The construction of
the workload matrix is elaborated further in Section 3. After that,
the mechanism searches for a \emph{strategy matrix} $A$ of size
$r$$\times$$n$, where $r$ is a positive integer. Intuitively, $A$
corresponds to another set of linear queries, such that every query
in $W$ can be expressed as a linear combination of the queries in
$A$. The matrix mechanism then answers the queries in $A$ under
differential privacy, and subsequently uses their noisy results to
answer queries in $W$.

The main challenge for applying the matrix mechanism to practical workloads is to identify an appropriate
strategy matrix $A$. Ref. \cite{LHR+10} provides two algorithms for this purpose. The first, based on iteratively solving a pair of related semidefinite programs, incurs $\mathcal{O}(m^3n^3)$ computational overhead, which is prohibitively expensive even for moderately large values of $m$ and $n$. The second solution computes an $\mathcal{L}_2$ approximation of the optimal strategy matrix $A$. This method, though faster than the first one, still incurs high costs as we show in the experiments. Further, the $\mathcal{L}_2$ approximation of the optimal strategy matrix often has poor quality. In fact, throughout our
experimental evaluations, we have never found a single setting where
this method obtains lower overall error
than the naive solution NOD that injects noise directly into the unit
counts. Although the matrix mechanism makes a significant theoretical contribution, so far its practice use is limited due to the lack of an effective implementation.
}


\section{Preliminaries}\label{sec:pre}

In this paper, we assume there are $n$ records in a database $D$,
i.e., $D=\{x_1,x_2,\ldots,x_n\}$. Each $x_i$ in $D$ is a real
number. To facilitate matrix manipulations, in the rest of the paper
we use a vector of size $n\times 1$ to denote the database,
i.e. $\{x_1,x_2,\ldots,x_n\}^T$. In Figure \ref{fig:intro-example},
for example, each record contains the number of HIV+ patients in a state of the USA. A query set $Q$ of cardinality $m$ is a mapping from the
database domain to real numbers, i.e., $Q:\mathbb{D}\mapsto
\mathbb{R}^m$.

\subsection{Differential Privacy}

A query processing mechanism $M$ is a randomized mapping from $\mathbb{D}\times
\mathbb{Q}$ to $\mathbb{R}^m$. Given an arbitrary query set $Q\in\mathbb{Q}$ and a database $D\in\mathbb{D}$, the mechanism $M$
returns a distribution on the query output domain $\mathbb{R}^m$.
Two databases $D_1$ and $D_2$ are neighbor databases \emph{iff} they differ on exactly one record, i.e.,
$D_1=\{x_1,x_2,\ldots,x_i,\ldots,x_n\}$ and
$D_2=\{x_1,x_2,\ldots,x'_i,\ldots,x_n\}$. A randomized mechanism
$M$ satisfies $\epsilon$-differential privacy if for every pair of
neighbor databases $D_1$ and $D_2$, we have
\begin{equation}
\forall Q\forall R:~ \Pr(M(Q,D_1)= R)\leq e^\epsilon\Pr(M(Q,D_2)= R)
\end{equation}

The above inequality implies that the mechanism $M$ always returns similar results on neighbor databases. This limits the adversary's confidence in inferring any record from the output of $M$, even when he or she knows all remaining records in the database.

In \cite{DMNS06}, Dwork et al. presented a general protocol to implement $\epsilon$-differential privacy, utilizing the concept of \emph{sensitivity}. Given a query set $Q\in\mathbb{Q}$, the sensitivity $\Delta$ is the maximal $\mathcal{L}_1$ distance between the exact query results on any neighbor databases $D_1$ and $D_2$, i.e.
\begin{equation}
\Delta=\max_{D_1,D_2}\|Q(D_1),Q(D_2)\|_1
\end{equation}

We emphasize that $\Delta$ only depends on the data domain $\mathbb{D}$ and the query set $Q$, not the actual data. Therefore, we simply assume such a constant $\Delta$ is public knowledge to everyone, including the adversary. The \emph{Laplace Mechanism} \cite{DMNS06}, $M_L$, outputs a randomized result $R$ on database $D$, following a Laplace distribution with mean $Q(D)$ and magnitude $\frac{\Delta}{\epsilon}$, i.e.,
\begin{equation}
\Pr(M_L(Q,D)=R)\propto \exp\left(\frac{\epsilon}{\Delta}\|R-Q(D)\|_1\right)
\end{equation}

This is equivalent to adding $m$-dimensional independent Laplace noise, as $Q(D)+Lap\left(\frac{\Delta}{\epsilon}\right)^m$, in which $Lap\left(\frac{\Delta}{\epsilon}\right)$ is a random variable following a zero-mean Laplace distribution with scale $\frac{\Delta}{\epsilon}$. Based on the definition of the Laplace mechanism, the expected squared error of the randomized query answer is $\frac{2m\Delta^2}{\epsilon^2}$, since the variance of $Lap(s)$ is $2s^2$ for any scale $s$. Note that the amount of error only depends on the sensitivity of the queries, regardless of the records in database $D$.

\subsection{Batch Linear Queries}\label{sec:pre:query}

As mentioned in the introduction, we focus on non-interactive linear queries in this paper. A linear query $q(D)$ is in the form of a linear function over the records in the database. Given a weight vector $\{w_1,w_2,\ldots,w_n\}^T$ of size $n$, the linear query returns the dot product between the weight vector and database vector, i.e.,
$$q(D)=w_1x_1+w_2x_2+\ldots+w_nx_n$$

We assume a batch of $m$ linear queries, $Q=\{q_1,q_2,\ldots,q_m\}$,
is submitted to the database at the same time. The query set $Q$
is thus represented by a \emph{workload matrix} $W$ with $m$ rows
and $n$ columns. Each entry $W_{ij}$ in $W$ is the $j$-th coefficient
for query $q_i$ on record $x_j$. Using the vector representation of
the database, i.e. $D=(x_1,x_2,\ldots,x_n)^T$, the query batch $Q$
can be exactly answered by calculating:
$$Q(D)=W D=\left(\sum_j W_{1j}x_j,\ldots,\sum_j W_{mj}x_j\right)^T$$

Based on the Laplace mechanism, two baseline solutions to enforce $\epsilon$-differential privacy on a query batch with workload
$W$ are as follows.

\noindent\textbf{Noise on data: } This solution, denoted as $M_D$, adds noise to
the original data. Given database $D$, $M_D$ generates a noisy database $D'$ using the
Laplace mechanism, i.e.,
$D'=D+Lap\left(\frac{\Delta}{\epsilon}\right)^n$. The query batch
$Q$ is then answered by replacing $D$ with $D'$. The whole mechanism
can be written in the form of manipulation on random variables, as follows.
\begin{equation}
M_D(Q,D)=W D'=W\left(D+Lap\left(\frac{\Delta}{\epsilon}\right)^n\right)
\end{equation}

Based on the linearity of expectation, it is straightforward to
calculate the expected squared error on the output,
$\frac{2\Delta^2}{\epsilon^2}\sum_{i,j}W_{ij}^2$, which is
proportional to the squared sum of the entries in $W$.

\noindent\textbf{Noise on results: } This baseline solution, denoted as $M_R$, adds noise to the query results instead of the original data.
Since the queries are linear queries, the sensitivity of the
query set is $\Delta'=\max_j \sum_i
|W_{ij}|\Delta$, i.e., the highest column absolute sum \cite{LHR+10}. Thus, $M_R$ outputs the following random
results.
\begin{equation}\label{eqn:m_r}
M_R(Q,D)= W D+Lap\left(\frac{\Delta'}{\epsilon}\right)^m
\end{equation}

Similarly, the expected squared error of the mechanism on query $Q$
is $2m\Delta'^2\epsilon^{-2}=2m\max_j \sum_i
W^2_{ij}\Delta^2\epsilon^{-2}$. By comparing their expected squared errors, we derive that $M_R$ outperforms $M_D$
by expectation, \emph{iff} $m\max_j\sum_i
W^2_{ij}<\sum_{j}\sum_i W^2_{ij}$. When $m\geq n$, this inequality
can never hold, implying that $M_R$ is more effective only when $m$
is smaller than $n$.


\subsection{Low Rank Matrices}

For any square matrix $A=\{A_{ij}\}$ of size $n\times n$, the trace
of the matrix is the sum of the diagonal entries in $A$, i.e.,
$\mbox{tr}(A)=\sum_i A_{ii}$. Given a matrix $W=\{W_{ij}\}$ of size
$m\times n$, the Frobenius norm of $W$ is the square root of the squared sum
over all entries, i.e., $\|W\|_F=\sqrt{\sum_{ij}(W_{ij})^2}$.
Following common notation, $W^T$ denotes the transposed
matrix of $W$.

Singular value decomposition (SVD) applies to any real-valued matrix $W$. Specifically, the
result of SVD on $W$ includes three matrices, $U$, $\Sigma$ and $V$,
such that $W=U\Sigma V$. Here, $U$, $\Sigma$, and $V$ are of size
$m\times s$, $s\times s$, and $s\times n$ respectively, where $m$ and $n$ are the number of rows and columns in $W$ respectively, and $s$ is
a positive integer no larger than $\min\{m,n\}$. Moreover, $U$ and
$V$ are row-wise and column-wise orthogonal matrices respectively.
$\Sigma$ is a diagonal matrix, which contains non-negative real
numbers on the diagonal and zeros in all the other entries. These diagonal entries,
$\{\lambda_1,\lambda_2,\ldots,\lambda_s\}$, are called eigenvalues of
the matrix $W$. The number of non-negative eigenvalues is called the rank of $W$, denoted as $rank(W)$.

When the rows and columns in the matrix $W$ are correlated, the rank of the matrix $W$ can be smaller than $m$ and $n$. In such cases, we say that $W$ is a low rank matrix. For example, when a group of records tend to appear together in a query, the workload matrix $W$ often exhibits strong column correlations. Similarly, when one query can be expressed as the linear combination of other queries, $W$ has strong row correlations. Both cases can be exploited to reduce the noise level necessary to satisfy differential privacy, as we showed in Section 1. Next we present the Low Rank Mechanism, a general solution to enforce differential privacy on a batch of linear queries, which utilizes the low rank property of the workload matrix to reduce noise.

\section{Workload Decomposition}\label{sec:formulation}

In this section, we propose a general workload matrix decomposition
technique that minimizes the error for a batch of linear queries. Recall that
the example in Figure \ref{fig:intro-example} shows that instead of adding
noise to the original data or query results (i.e., methods NOD and NOR),
it is sometimes possible to construct another linear basis that leads to higher overall query accuracy. To build such a basis, we
partition the workload matrix $W$ into the product of two components, $B=\{B_{ij}\}$
of size $m\times r$ and $L=\{L_{jk}\}$ of size $r\times n$, such
that $W=BL$. Note that $r$ can be larger than the rank of the
workload matrix $W$. Given the matrix decomposition, we design
general mechanism for adding noise to $LD$ ($D$ is the dataset), and analyze the expected
squared error. We first formally define the concepts of \emph{query scale} and \emph{query sensitivity}, for a given
decomposition $W=BL$.

\begin{definition}\label{def:scale}\textbf{Query Scale}\\
Given a workload decomposition $W=BL$, the scale of the
decomposition, denoted by $\Phi(B,L)$, is the squared sum of the
entries in $B$, i.e., $\Phi(B,L)=\sum_{i,j}B_{ij}^2$.
\end{definition}

\begin{definition}\label{def:sensitivity}\textbf{Query Sensitivity}\\
Given a workload decomposition $W=BL$, the sensitivity of the
decomposition, denoted by $\Delta(B,L)$, is the maximal absolute sum
of any column in $L$, i.e., $\Delta(B,L)=\max_j\sum_i |L_{ij}|$.
\end{definition}

Since $W=BL$, the linear query batch can be answered by calculating $Q(D)=WD=BLD$. Unlike solutions NOD and NOR, we inject noise into the intermediate result $LD$ to enforce differential privacy. Since $LD$ is another group of linear queries, we can apply NOR on $Q'(D)=LD$ with Eq. (\ref{eqn:m_r}). The sensitivity of the new linear query batch is $\Delta(B,L)$, which leads to the following differential privacy mechanism $M_P(Q,D)$ with respect to the workload decomposition $W=BL$.
\begin{equation}\label{eqn:part_mech}
M_P(Q,D)=B\left(LD+Lap\left(\frac{\Delta(B,L)}{\epsilon}\right)^r\right)
\end{equation}

The error analysis of $M_P(Q,D)$ is complicated as its adds noise at an intermediate step. The following lemma shows that the error is linear in the query scale, and quadratic in the query sensitivity.
\begin{lemma}\label{lem:decomp_error}
The expected squared error of $M_P(Q,D)$ with respect to the
decomposition $W=BL$ is
$2\Phi(B,L)\left(\Delta(B,L)\right)^2/\epsilon^2$.
\end{lemma}

Accordingly, we reduce the problem to finding the optimal workload decomposition $W=BL$ that minimizes $\Phi(B,L)\left(\Delta(B,L)\right)^2$. However, this optimization problem is difficult to solve, since the objective function is the product of $\Phi(B,L)$ and $\Delta(B,L)$, and $\Delta(B,L)$ may not be derivable. To address this problem, we first prove an interesting property of the workload decomposition, which implies that the exact query sensitivity is actually not important.

\begin{lemma}\label{lem:rescale}
Given a workload decomposition $W=BL$ and a positive constant $\alpha$, we can always construct another decomposition $W=B'L'$ such that $B'=\alpha B$ and $L'=\alpha^{-1}L$, satisfying
$$\Phi(B,L)\left(\Delta(B,L)\right)^2=\Phi(B',L')\left(\Delta(B',L')\right)^2$$
\end{lemma}

According to the above lemma, the balance between scale and sensitivity is not important, as we can always build another equivalent workload decomposition with arbitrary sensitivity. This motivates us to formulate a new optimization program, which focuses on minimizing the query scale while fixing the query sensitivity. The following theorem formalizes this claim.

\begin{theorem}\label{the:main}
Given the workload $W$, $W=BL$ is the optimal workload decomposition to minimize expected squared error if $(B,L)$ is the optimal solution to the following program:
\end{theorem}
\begin{equation}\label{eqn:opt-problem}
\begin{split}
\mbox{Minimize: } \mbox{tr}(B^TB)\\
\mbox{s.t.~~}  W=BL\\
\forall j \sum_i |L_{ij}|\leq 1
\end{split}
\end{equation}

In the optimization problem above, we are allowed to specify the number of columns in the matrix $B$, i.e. the rank $r$ of the matrix product $BL$. This enables us to generate matrices of significantly lower rank than the strategy matrix proposed in \cite{LHR+10}. We thus use \emph{Low Rank Mechanism} to denote the general query processing scheme in Eq. (\ref{eqn:part_mech}), using the optimal decomposition solution to Formula (\ref{eqn:opt-problem}).

\subsection{Optimality Analysis}\label{sec:analysis}

In this subsection, we analyze the optimality of our optimization formulation. Specifically, we show that the utility of our proposed mechanism almost reaches the known utility lower bound for linear queries under differential privacy \cite{HT10}.
\vspace{-5pt}
\begin{lemma}\label{lem:upper}
Given a workload matrix $W$ of rank $r$ with eigenvalues $\{\lambda_1,\ldots,\lambda_r\}$, the expected squared error of $M_P(Q,D)$ w.r.t. the optimal decomposition $W=B^*L^*$ in low rank mechanism is
bounded above by $\sum^r_{k=1}\lambda^2_kr/\epsilon^2$.
\end{lemma}

Using the geometric analysis technique under orthogonal projection \cite{HT10}, the following lemma reveals a lower bound on the squared error for linear queries.

\begin{lemma}\label{lem:lower}Given a workload matrix $W$ of rank $r$ with eigenvalues $\{\lambda_1,\ldots,\lambda_r\}$, the expected squared error of any
$\epsilon$-differential privacy mechanism is at least
$$\Omega\left(\left(\frac{2^r}{r!}\prod^r_{k=1}\lambda_k\right)^{2/r}r^3/\epsilon^2\right)$$
\end{lemma}

Assume that all the eigenvalues $\{\lambda_1,\lambda_2,\ldots,\lambda_r\}$ of workload $W$ are
ordered in non-ascending order. We use $C=\lambda_1/\lambda_r$ to
denote the ratio between the largest eigenvalue and the smallest non-zero
eigenvalue. The following theorem discusses the tightness of low rank mechanism on error minimization. In particular, it proves the optimality of the result decomposition $W=B^*L^*$ with respect to Formula (\ref{eqn:opt-problem}).

\begin{theorem}\label{the:opt} When $r>5$, the mechanism $M_p{\left(Q,D\right)}$
using $W=B^*L^*$ is an $\mathcal{O}(C^2r)$-approximately optimal solution w.r.t. the set of all
non-interactive $\epsilon$-differential privacy mechanisms.
\end{theorem}

When $C$ is close to 1, all non-zero eigenvalues are close
to each other and the mechanism under our decomposition optimization
program outputs results that well approximate the lower bound. This result answers one of the questions in
\cite{HT10}, in which the authors discussed possible orthogonal
projections but did not provide a concrete algorithm to identify the optimal
projection. Our formulation can be regarded as an implementation of orthogonal projection with almost constant
approximation. Therefore, our result fills the gap between theory and practice.

\subsection{Relaxation on Decomposition}

Theorem \ref{the:opt} shows that our decomposition leads to results with a tight bound. However, when there are very small eigenvalues in the workload matrix $W$, the bound in the theorem becomes loose. On the other hand, these small eigenvalues contribute little to the workload matrix $W$. This observation motivates us to design
a new optimization formulation, in which $BL$ does not necessarily match $W$, but within a small error tolerance. This enables the
formulation to find a more compact decomposition, such that the $r$ used in $B$ and $L$ can be smaller than the actual rank of $W$.

To do this, we introduce a new parameter $\gamma$ to bound the
difference between $W$ and $BL$ in terms of the Frobenius norm. This leads to
a new optimization problem:
\begin{equation}\label{eqn:relaxed-problem}
\begin{split}
\mbox{Minimize: } \mbox{tr}(B^TB)\\
\mbox{s.t.~~}  \|W-BL\|_F\leq\gamma\\
\forall j \sum_i |L_{ij}|\leq 1
\end{split}
\end{equation}

After finding the optimal $(B,L)$  for the problem in Formula \ref{eqn:relaxed-problem}, the mechanism $M_P(Q,D)$ outputs query results using Eq. (\ref{eqn:part_mech}). The error of this new mechanism is also bounded, as stated in the following theorem.

\begin{theorem}\label{the:new_error}
The expected squared error of $M_P(Q,D)$ using the decomposition $(B,L)$ satisfying Eq. (\ref{eqn:relaxed-problem}) is at most
$$2\mbox{tr}(B^TB)/\epsilon^2+\gamma\sum_{i}x^2_{i}$$
\end{theorem}

While Theorem \ref{the:new_error} implies the possibility of estimating the
optimal $\gamma$, it is not practical to implement it directly, because
this estimation depends on the data, i.e., $\sum_{i}x^2_{i}$. In our
experiments, we test different values of $\gamma$, and report their relative performance, regardless of the data distribution.

\section{Decomposition Algorithm}\label{sec:opt}

The previous section formulates the workload matrix decomposition problem as an optimization program, which is rather complicated and non-trivial to solve. This section describes an effective and efficient solution for this program, based on the inexact Augmented Lagrangian Multiplier (ALM) method \cite{Conn1997,Lin2010}.

The main challenge in solving the optimization program of Formula (\ref{eqn:relaxed-problem}) is the non-smooth $\mathcal{L}_1$ regularized term. The projected gradient method \cite{Duchi2008} is considered one of the most efficient general algorithms to solve these problems. Following the strategy used in \cite{Conn1997}, we treat the $\mathcal{L}_1$ regularized term separately and approximately minimize a sequence of Lagrangian subproblems. Our inexact Augmented Lagrangian method for workload matrix decomposition problem is summarized in Algorithm \ref{algorithm:Lagrange}.

In order to handle the linear constraints
$\|W-BL\|_F\leq\gamma\rightarrow0$, in which $W \in \mathbb{R}^{m
\times n}$, $B \in \mathbb{R}^{m \times r}$ and  $L \in
\mathbb{R}^{r \times n}$, the inexact Augmented Lagrangian method
introduces a positive penalty item $\beta \in \mathbb{R}$ and the
Lagrange multiplier $\pi \in \mathbb{R}^{m \times n}$. The update on
$\beta$ and $\pi$ follows the standard strategy used in
\cite{Conn1997,Lin2010}. Given fixed $\beta$ and $\pi$ in each
iteration, the algorithm aims to find a pair of new $B$ and $L$ to
minimize the following subproblem:


\vspace{-10pt}

\begin{eqnarray}
\mathcal{J}(B,L,\beta,\pi)  =  \frac{1}{2}\mbox{tr}(B^TB)+
\langle\pi,W-BL \rangle \nonumber  + \frac{\beta}{2}
\|W-BL\|_F^2\nonumber \\ ~~s.t.~~ \forall j \sum_i |L_{ij}|\leq
1\nonumber
\end{eqnarray}

\begin{algorithm}[t]
\caption{\label{algorithm:Lagrange} {\bf Workload Matrix
Decomposition}}
\begin{algorithmic}[1]
\STATE  Initialize $\pi^{(0)}=\mathbf{0}\in \mathbb{R}^{m\times n},
\beta^{(0)}=1, k=1$
  \label{OutIter}\WHILE{not converged}
  \STATE//Approximately solve the subproblem
  \label{InIter}\WHILE {not converged}\label{main_alg:lag:begin}
  \STATE \label{step1} $B^{(k)} \leftarrow$ update $B$ using Eq. (\ref{UpdateB})
  \STATE \label{step2} $L^{(k)} \leftarrow$ run Algo. \ref{algorithm:Nesterov} to update $L$ w.r.t. Formula (\ref{eqn:Lagrangian-Seq-L})
  \ENDWHILE\label{main_alg:lag:end}
  \STATE Compute $\tau = \|W-B^{(k)}L^{(k)}\|_F$
  \IF{$\tau$ is sufficiently small or $\beta$ is sufficiently large}
  \STATE return $B^{(k)}$ and $L^{(k)}$
  \ENDIF
  \IF{$k$ is divisible by 10}
  \STATE $\beta^{(k+1)} = 2\beta^{(k)}$
  \ENDIF
  \STATE $\pi^{(k+1)} = \pi^{(k)} + \beta^{(k+1)} \left(W-B^{(k)}L^{(k)}\right)$
  \STATE $k=k+1$
  \ENDWHILE
\end{algorithmic}
\end{algorithm}

This is a Bi-Convex optimization problem, which can be solved by block gradient descent via
alternately optimizing $B$ and $L$. Based on the formulation above,
optimizing $B$ is straightforward. Since the gradient with respect
to $B$ can be computed as:

\vspace{-5pt}
$$\frac{\partial \mathcal{J}}{\partial B} = B - \pi L^T + \beta BLL^T
- \beta WL^T$$,

\noindent based on the fact that $\mathcal{J}(\cdot)$ is convex
with respect to $B$, we can set $\frac{\partial
\mathcal{J}}{\partial B}=0$, and obtain a closed form solution to
update $B$:

\vspace{-5pt}

\begin{equation}\label{UpdateB}
B=\left(\beta W L^T+\pi L^T\right) \left(\beta LL^T + I\right)^{-1}
\end{equation}

The second step is to optimize $L$, which is equivalent to solving
the following quadratic programming problem:

\vspace{-10pt}
\begin{equation}\label{eqn:Lagrangian-Seq-L}
\begin{split}
\mathcal{G}(L)=\frac{\beta}{2} \mbox{tr}\left(L^TB^TBL\right) -  \mbox{tr}\left(\left(\beta W+\pi\right)^TBL\right)\\
    \mbox{s.t.~~}  \forall j \sum_i |L_{ij}|\leq 1
\end{split}
\end{equation}

In order to minimize Eq. (\ref{eqn:Lagrangian-Seq-L}) under constraints, we employ
Nesterov's first order optimal method \cite{Nesterov03} to
accelerate the gradient decent. Nesterov's method has a much faster
convergence rate than traditional methods such as the subgradient
method or the na\"{i}ve projected gradient descent. In particular, the
gradient of $\mathcal{G}(L)$ with respect to $L$ is

$$\frac{\partial \mathcal{G}}{\partial L} = \beta B^TBL-\beta B^TW - B^T \pi$$

$L$ is updated by gradient descent while ensuring that the
$\mathcal{L}_1$ regularized constraint on $L$ is satisfied. This can
be done by solving the following optimization problem:

\begin{equation}\label{eqn:L1Projection}
    \min_L \|L-L^{(t)}\|_F^2 , s.t. ~~\forall j \sum_i |L_{ij}|\leq 1,
\end{equation}

\noindent in which $L^{(t)}$ denotes the last feasible solution after exactly
$k$ iterations. Since Formula (\ref{eqn:L1Projection}) can be decoupled
into $r$ independent $\mathcal{L}_1$ regularized sub-problems, it
can be solved efficiently by $\mathcal{L}_1$ projection methods
\cite{{Duchi2008}}. The complete algorithm for the projection method
is summarized in Algorithm \ref{algorithm:Nesterov}.

%
%

\begin{algorithm}[t]
\caption{\label{algorithm:Nesterov} {\bf Nesterov's Projection
Gradient Method}}
\begin{algorithmic}[1]
\STATE  input: $\mathcal{G}(L), \frac{\partial \mathcal{G}}{\partial L}, L^{(0)}$ \\
\STATE $\chi=r\cdot n \cdot 10^{-12}$, Lipschitz parameter: $\omega^{(0)}=1$\\
 \STATE Initializations: $L^{(1)}=L^{(0)}, \delta^{(-1)}=0, \delta^{(0)}=1, t=1$
  \WHILE{not converged}
      \STATE $\alpha=\frac{\delta^{(t-2)}-1}{\delta^{(t-1)}}$, $S=L^{(t)}+\alpha(L^{(t)}-L^{(t-1)})$
      \FOR{$j=0$ to ...}
      \STATE $\omega=2^j \omega^{(t-1)}$, $U = S - \frac{1}{\omega} \nabla_{S}$
      \STATE Project $U$ to the feasible set to obtain $L^{(t)}$ (i.e. solve Formula (\ref{eqn:L1Projection}))
      \IF{ \label{stopping1} $\|S-L^{(t)}\|_F<\chi$}
        \STATE  return;
      \ENDIF
      \STATE  Define function: $\mathcal{J}_{\omega,S}(U)=\mathcal{G}(S)+\langle \frac{\partial G}{\partial U}, U-S\rangle+\frac{\omega}{2}\|U-S\|_F^2$
      \IF{$ \mathcal{G}(L^{(t)}) \leq \mathcal{J}_{\omega,S}(U)$}
        \STATE $\omega^{(t)}=\omega; L^{(t+1)}=L^{(t)}$; break;
      \ENDIF
      \ENDFOR
      \STATE Set $\delta^{(t)}= \frac {1+ \sqrt{{1+4(\delta^{(t-1)})^2}}}{2}$
      \STATE $t=t+1$
      \ENDWHILE
\STATE return $L^{(t)}$
\end{algorithmic}
\end{algorithm}

\begin{figure*}[t]
\centering \subfigure[`WDiscrete']
{\includegraphics[width=0.3\textwidth]{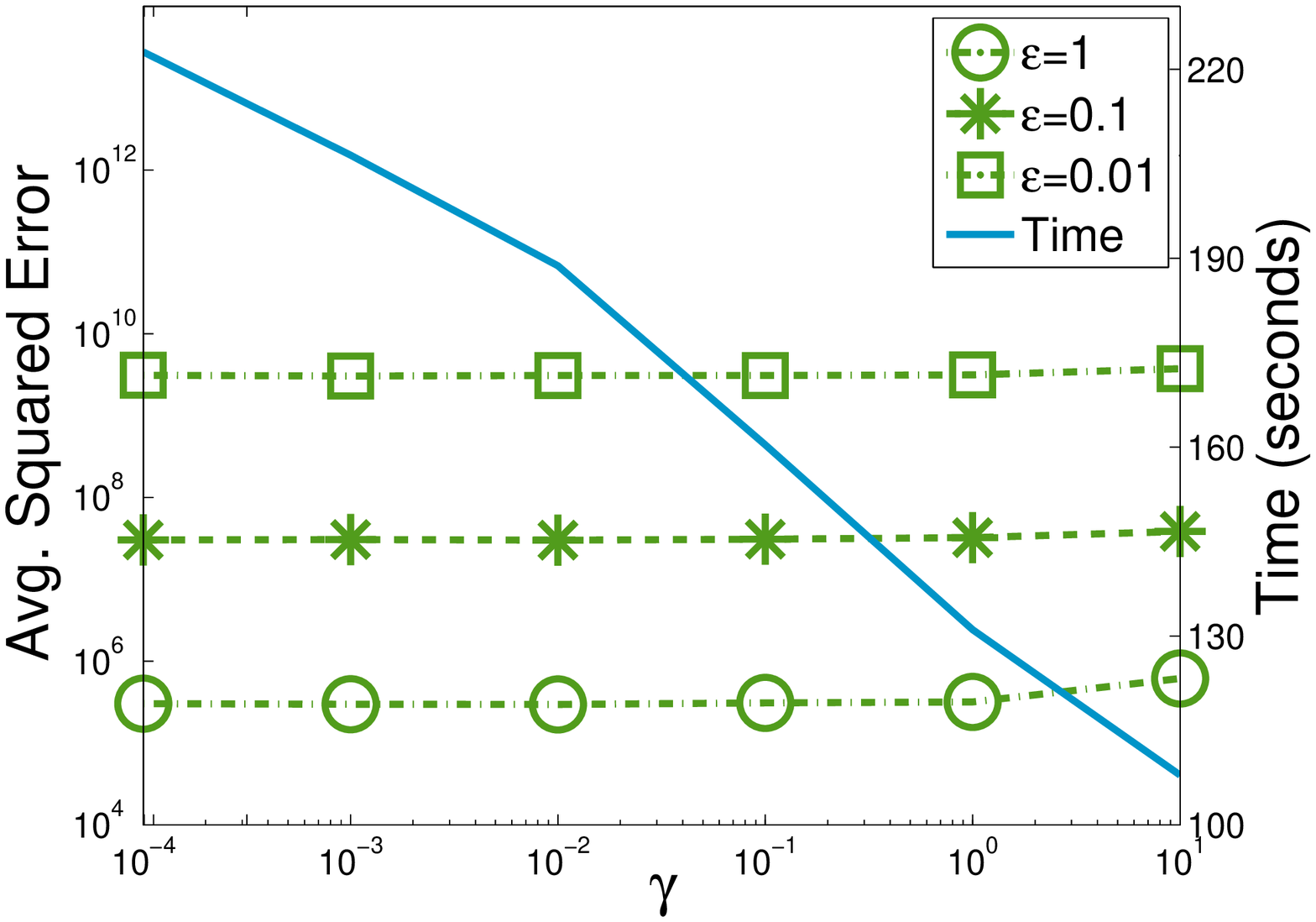}}
\subfigure[`WRange']
{\includegraphics[width=0.3\textwidth]{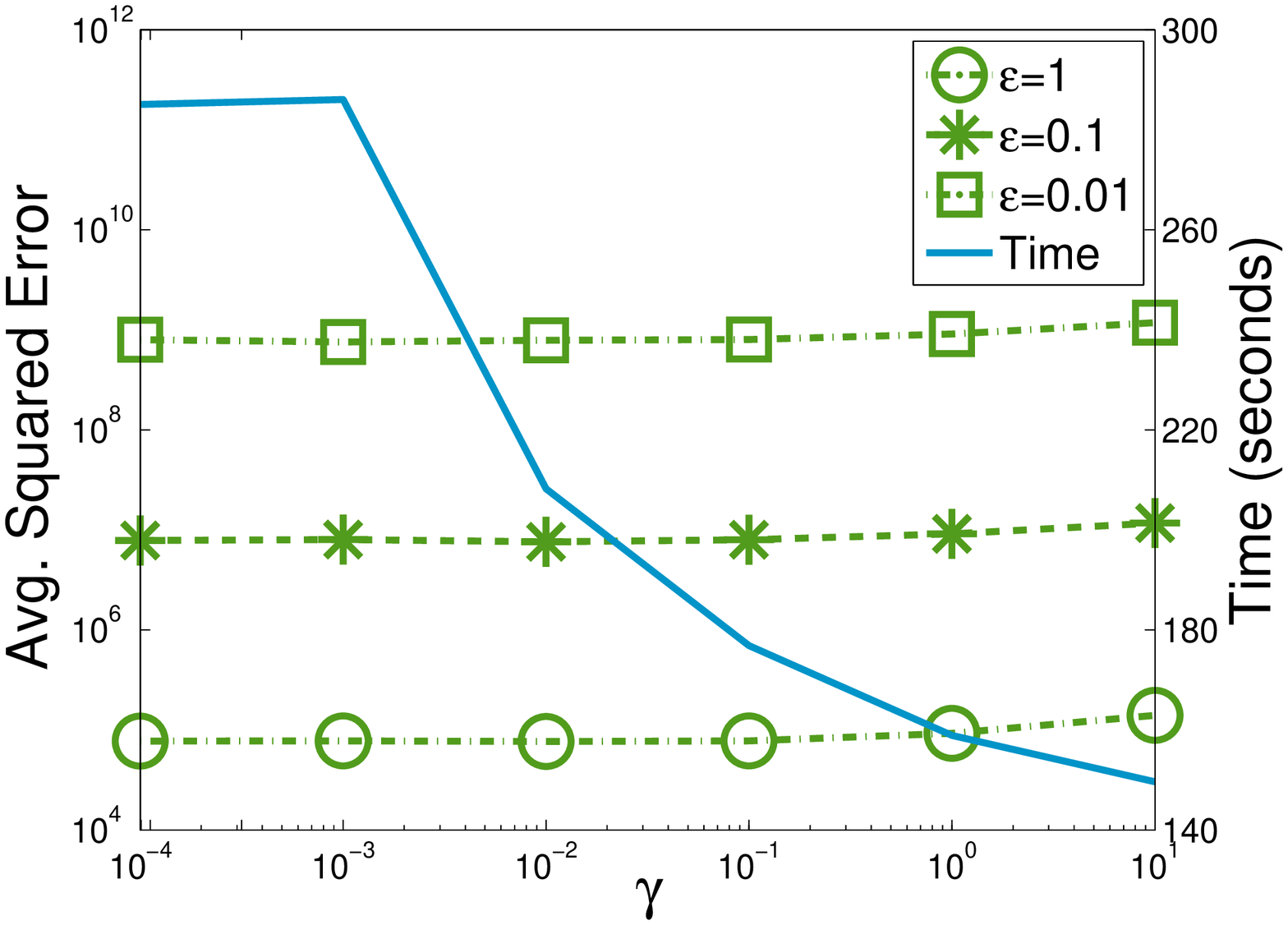}}
\centering \subfigure[`WRelated']
{\includegraphics[width=0.3\textwidth]{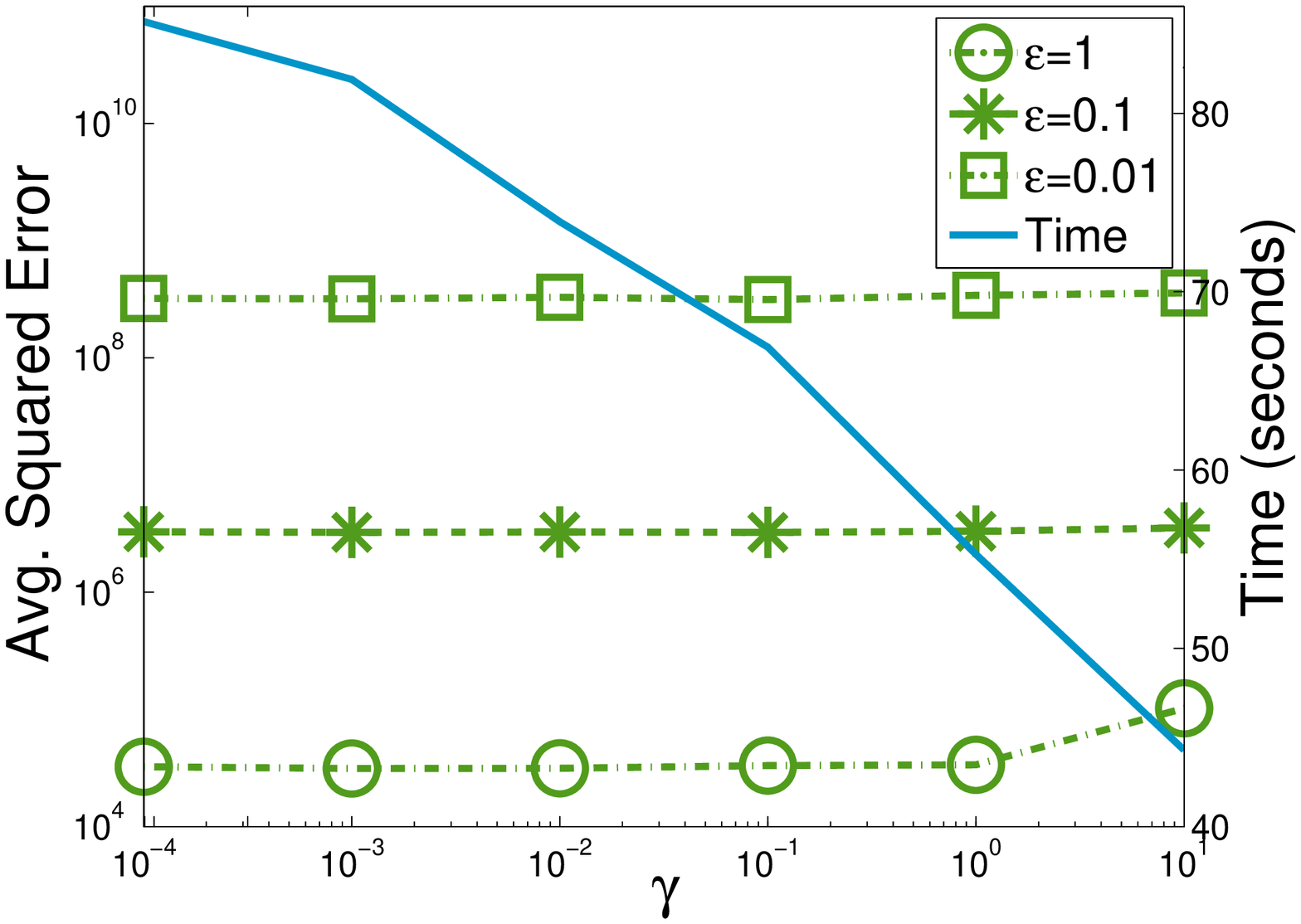}}
\vspace{-5pt}
\caption{Effect of varying relaxation parameter $\gamma$ with the \emph{Search Logs} dataset for LRM}
\label{fig:exp:gamma}
\vspace{-5pt}
\end{figure*}

\begin{figure*}[t]
\centering \subfigure[`WDiscrete']
{\includegraphics[width=0.3\textwidth]{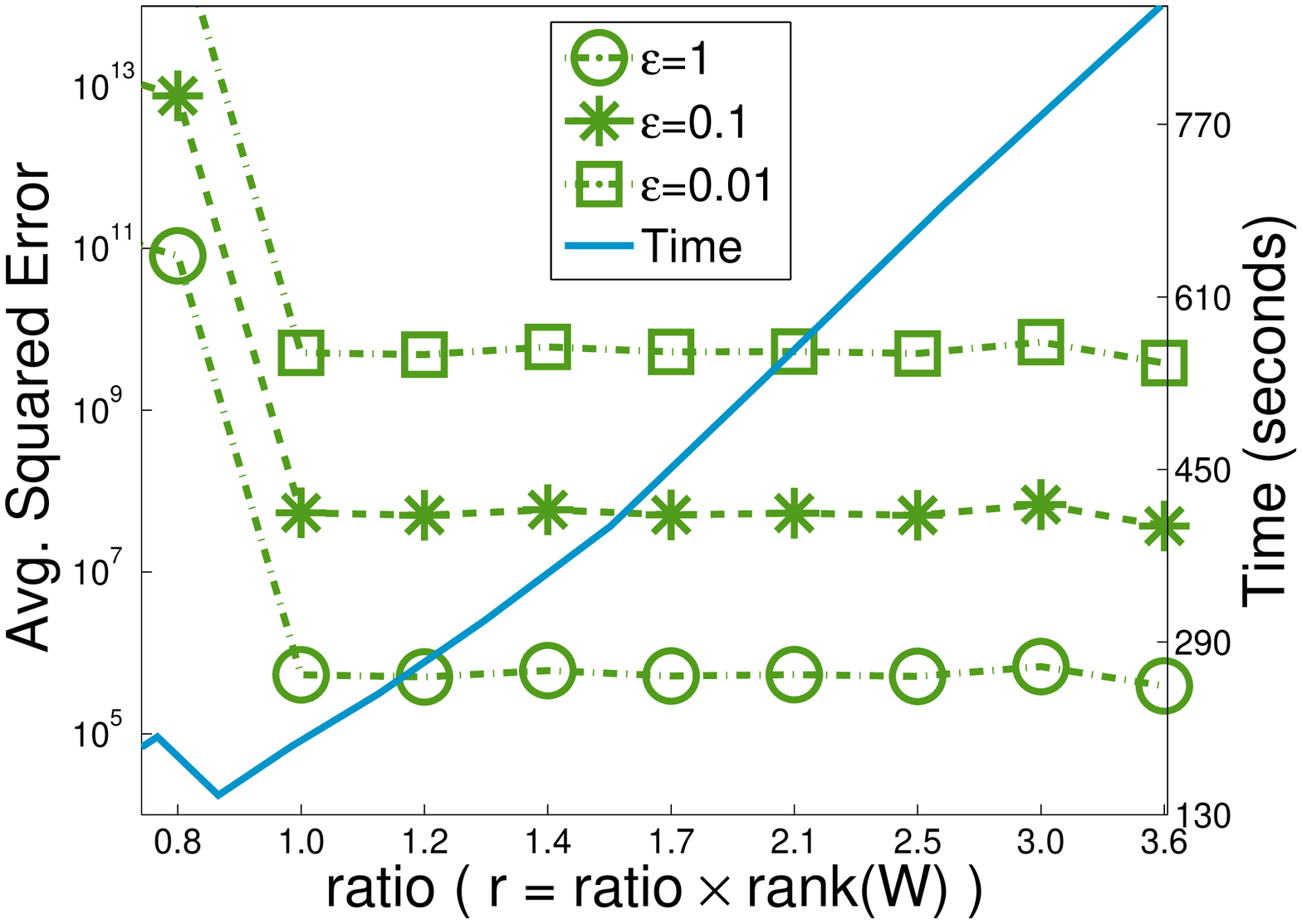}}
\subfigure[`WRange']
{\includegraphics[width=0.3\textwidth]{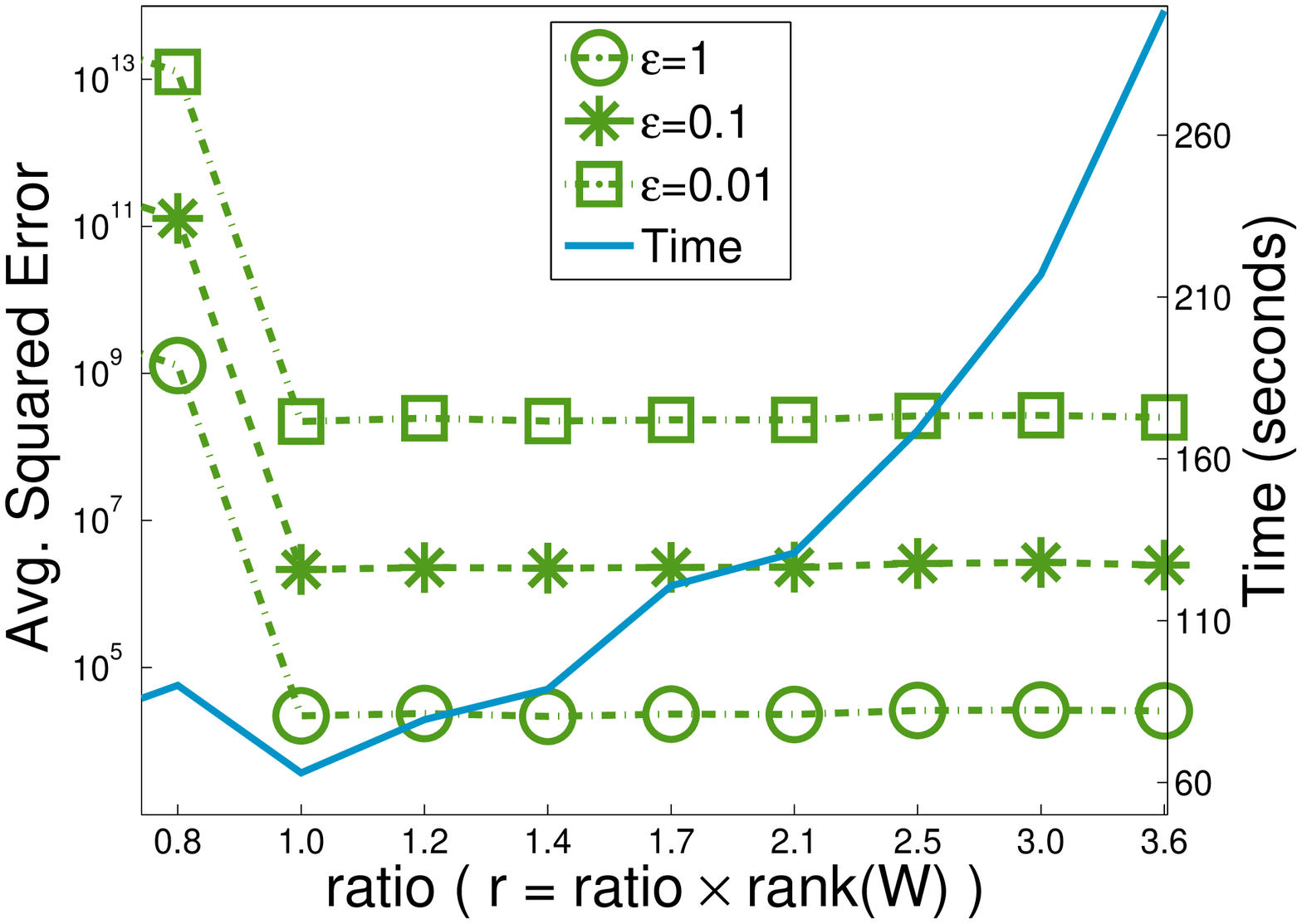}}
\centering \subfigure[`WRelated']
{\includegraphics[width=0.3\textwidth]{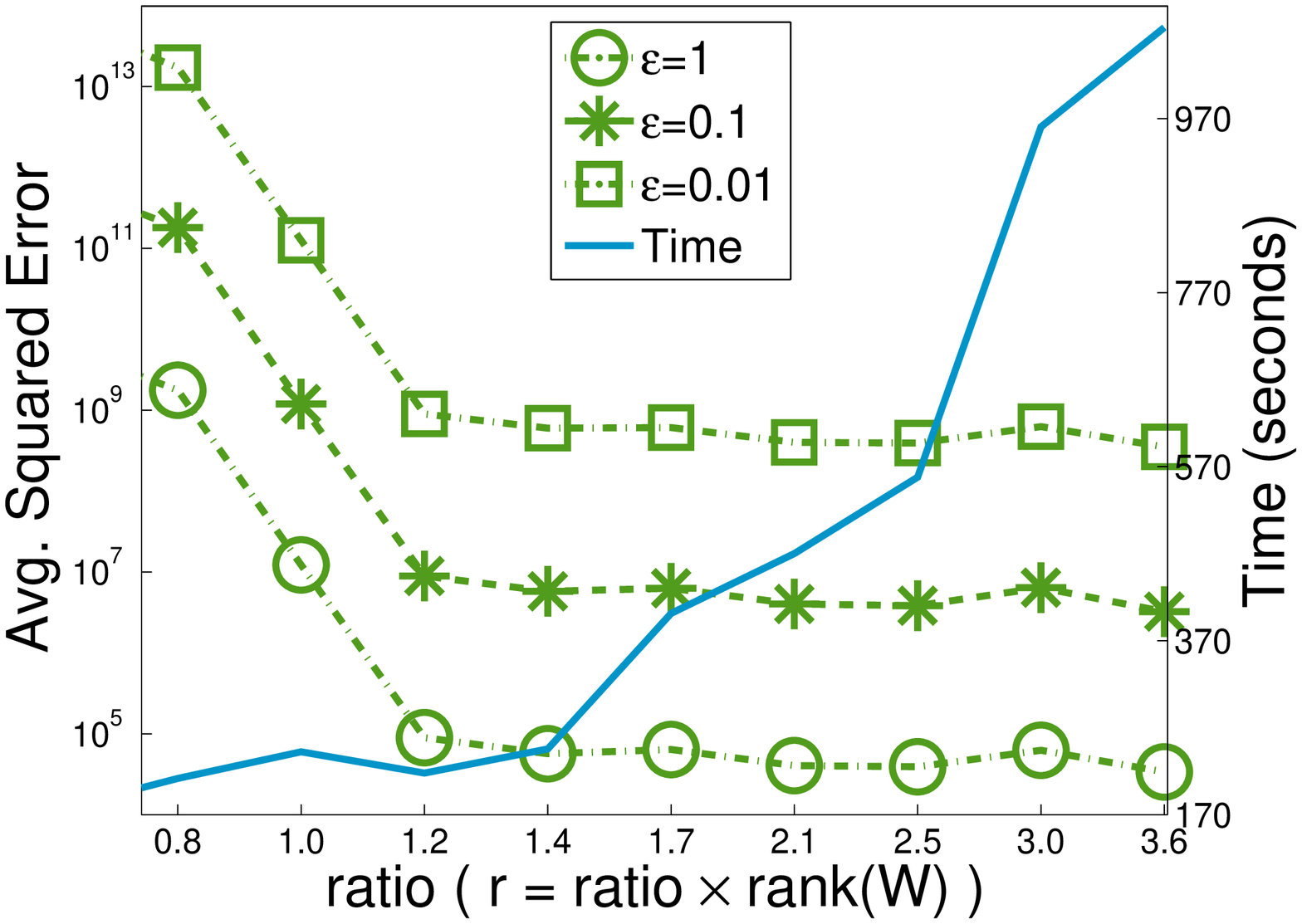}}
\vspace{-5pt}
\caption{Effect of varying $r$ with \emph{Search Logs} dataset for LRM}
\label{fig:exp:r}
\vspace{-5pt}
\end{figure*}

\begin{figure*}
\centering \subfigure[\emph{Search Logs}]
{\includegraphics[width=0.3\textwidth]{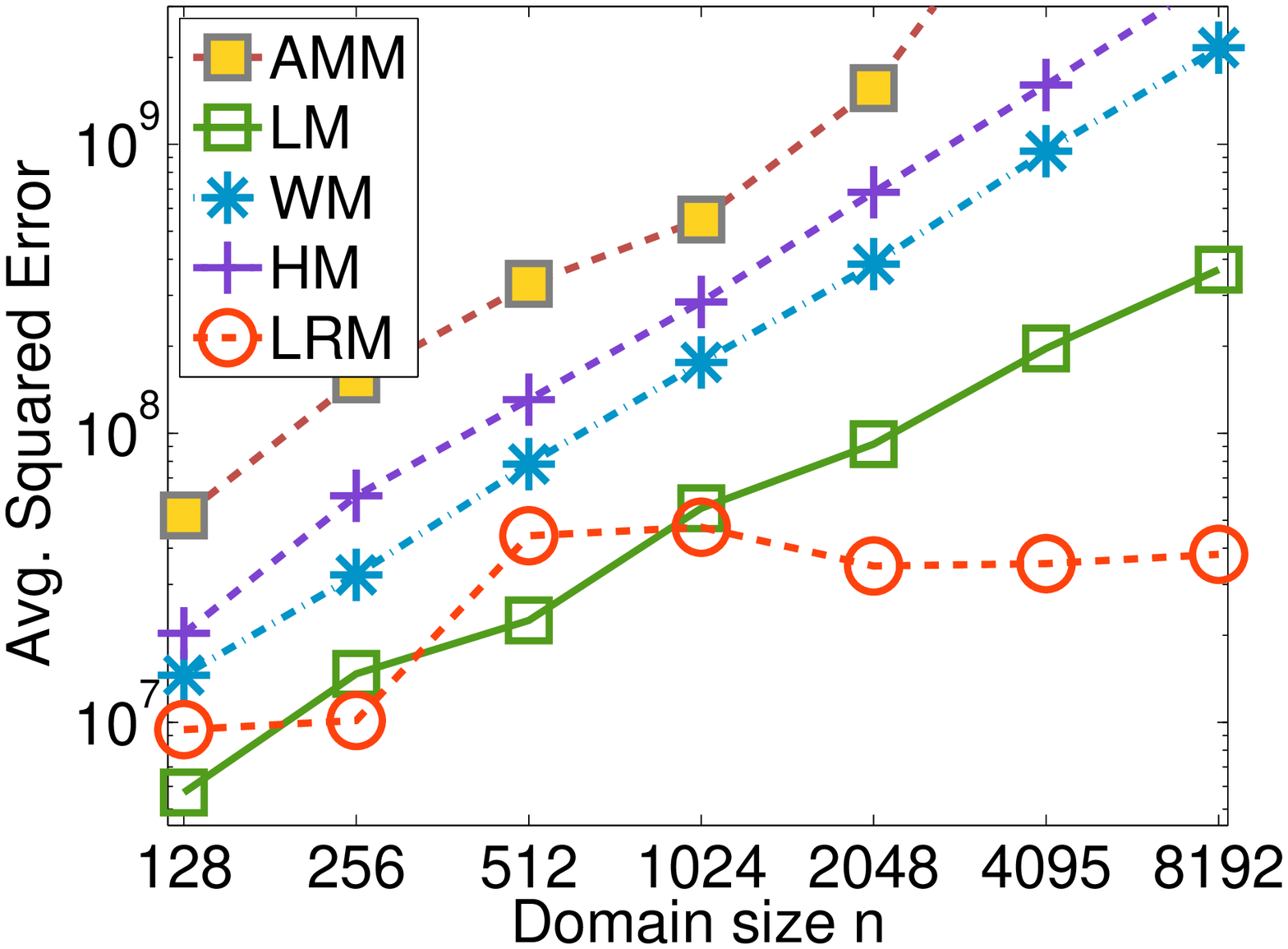}}
\subfigure[\emph{NetTrace}]
{\includegraphics[width=0.3\textwidth]{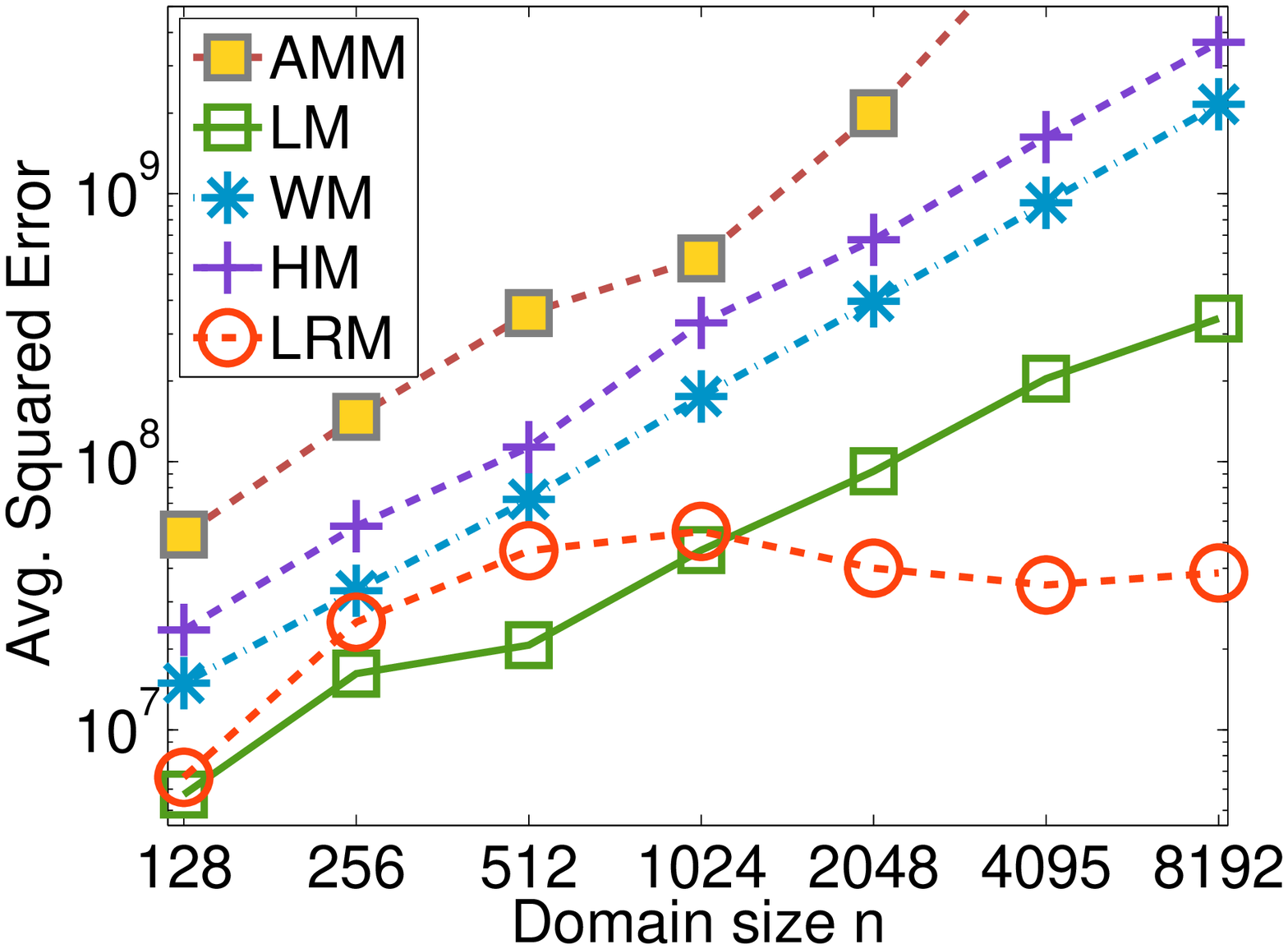}}
\centering \subfigure[\emph{Social Network}]
{\includegraphics[width=0.3\textwidth]{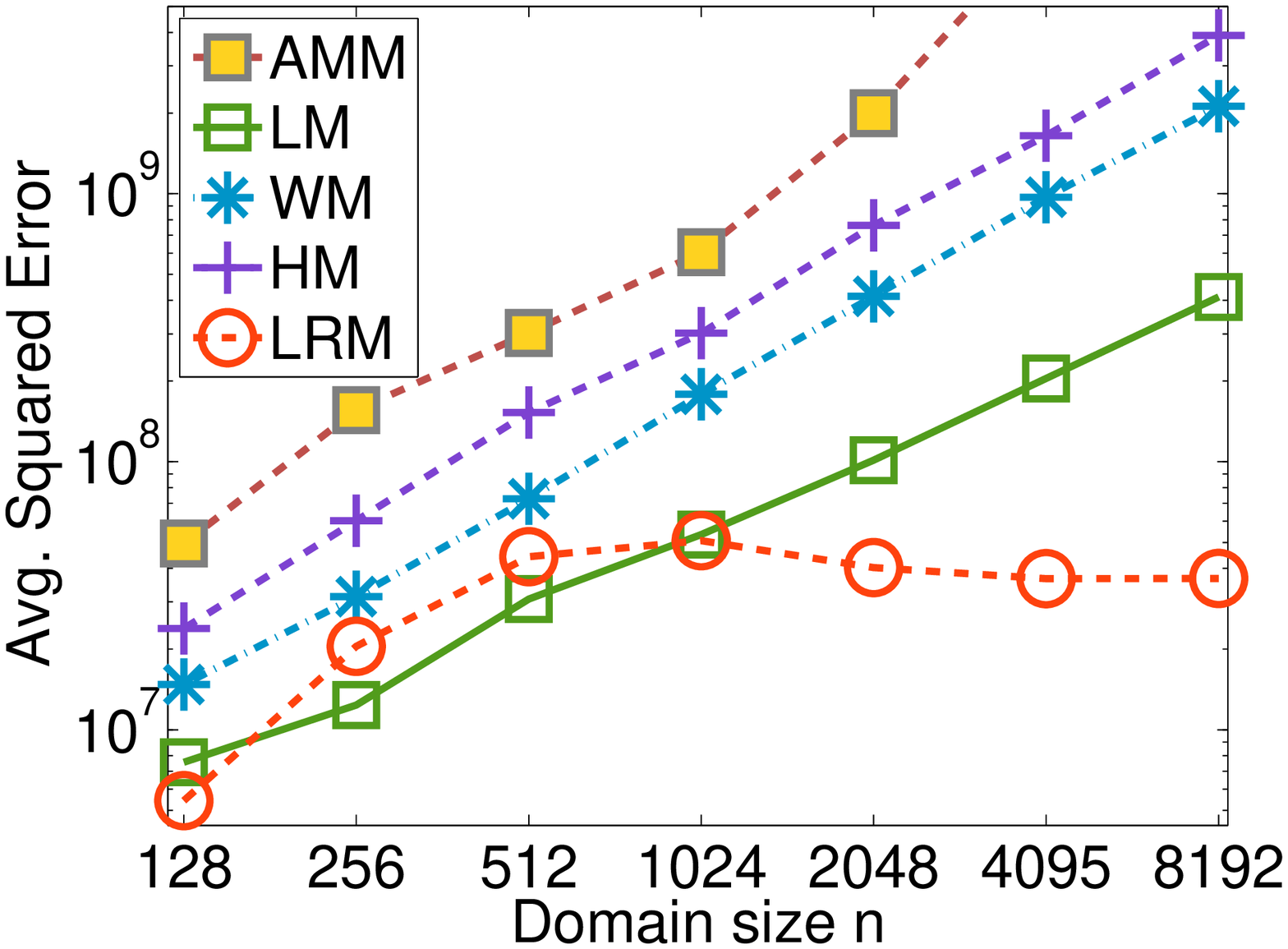}}
\vspace{-5pt}
\caption{Effect of varying domain size $n$ on workload `WDiscrete' with
$\epsilon=0.1$}\label{fig:exp:n:WDiscrete}
\vspace{-10pt}
\end{figure*}

\textbf{Convergence Analysis:} In each iteration, the algorithm
solves a sequence of Lagrangian subproblems by optimizing $B$ (step
\ref{step1}) and $L$ (step \ref{step2}) alternatingly.
The algorithm stops when a sufficiently small $\gamma$ is achieved or
the penalty parameter $\beta$ is sufficiently large. It suffices
to guarantee that $L$ converges to the optimal solution \cite{Lin2010}.
Although the objective function is non-smooth, the algorithm
possesses excellent convergence properties. To be precise, we formally
establish the following convergence statement.
\begin {theorem} \label{LowRankDPConvergence}
If $(B^{(k)},L^{(k)})$ is the temporary solution after the $k$-th iteration and $(B^*,L^*)$ is the optimal solution to Formula (\ref{eqn:opt-problem}), we have
\begin{equation}\nonumber
\left|\mbox{tr}(B^{(k)}B^{(k)}) -  \mbox{tr}(B^*B^*)\right|  \leq
\mathcal{O}\left(\frac{1}{\beta^{k-1}}\right)
\end{equation}
\end {theorem}

Since $\beta^{(k)}$ doubles after every 10 iterations, the algorithm
converges rapidly. This proves the fast convergence property of our
algorithm. 
%
 
\textbf{Complexity Analysis:} The total number of variables in $B$ and $L$ is $(m+n)r$.
Each update on $B$ in Eq.
(\ref{UpdateB}) takes $\mathcal{O}(r^2m)$ time, while each update on $L$ takes $\mathcal{O}(r^2 n)$ time.
If Algorithm\ref{algorithm:Lagrange} converges to a local minimum
with $N_{in}$ inner iterations (at line 4 in Algorithm \ref{algorithm:Lagrange}) and $N_{out}$
outer iterations (at line 2 in Algorithm \ref{algorithm:Lagrange}), the total complexity of
Algorithm \ref{algorithm:Lagrange} is $\mathcal{O}(N_{in} \times
N_{out}\times(r^2m + r^2n))$.

\begin{figure*}[t]
\centering \subfigure[\emph{Search Logs}]
{\includegraphics[width=0.3\textwidth]{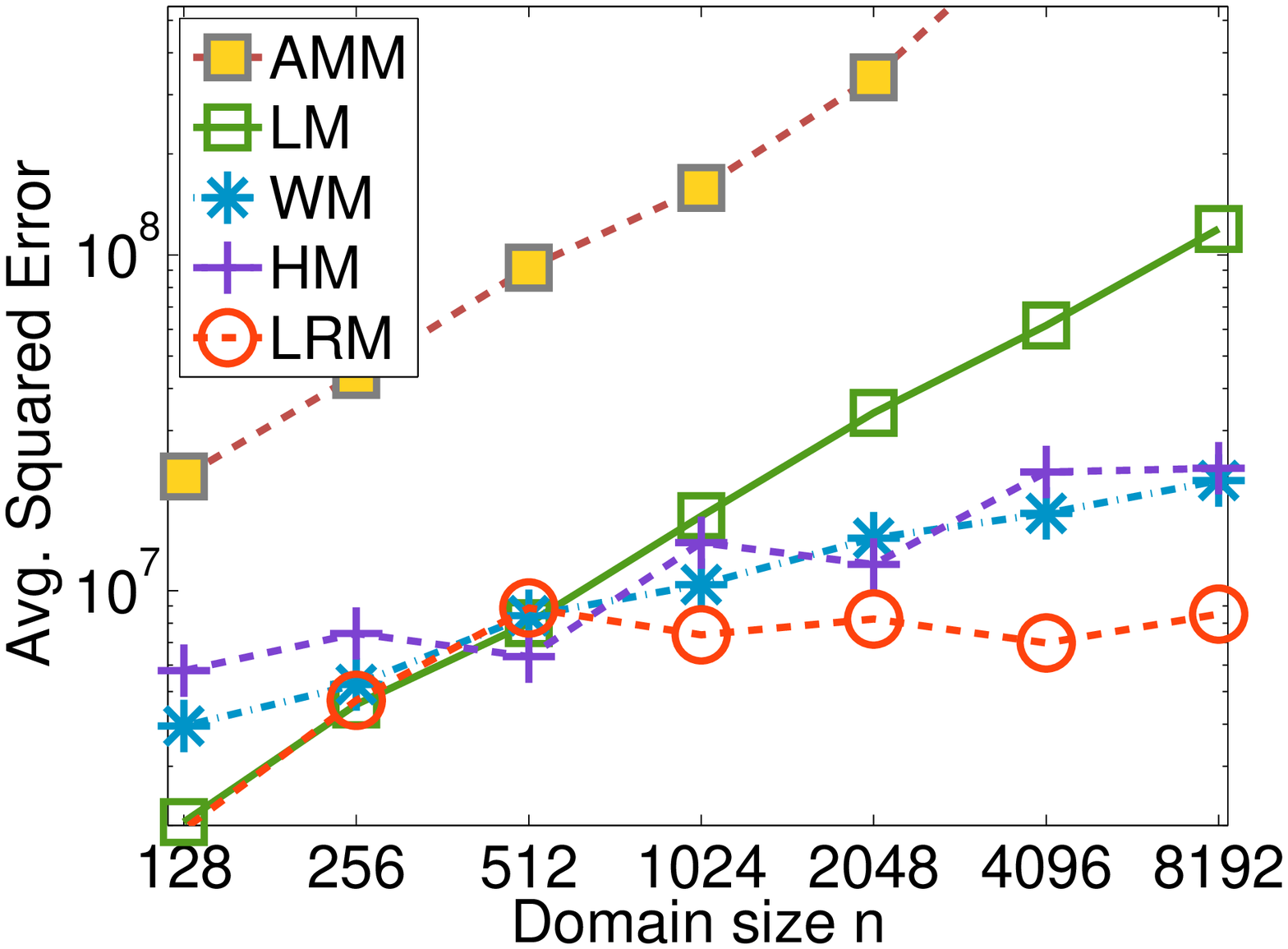}}
\subfigure[\emph{NetTrace}]
{\includegraphics[width=0.3\textwidth]{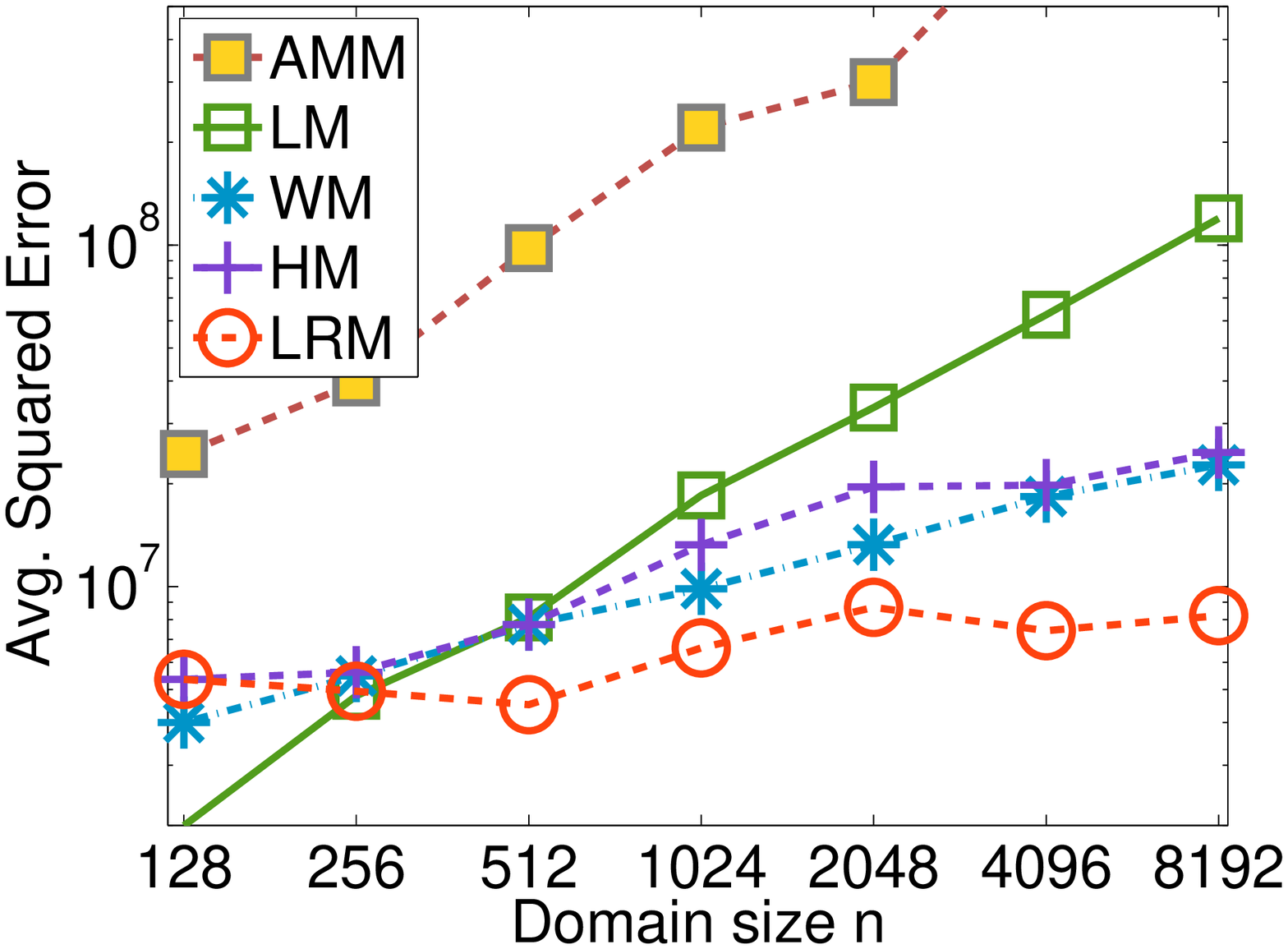}}
\centering \subfigure[\emph{Social Network}]
{\includegraphics[width=0.3\textwidth]{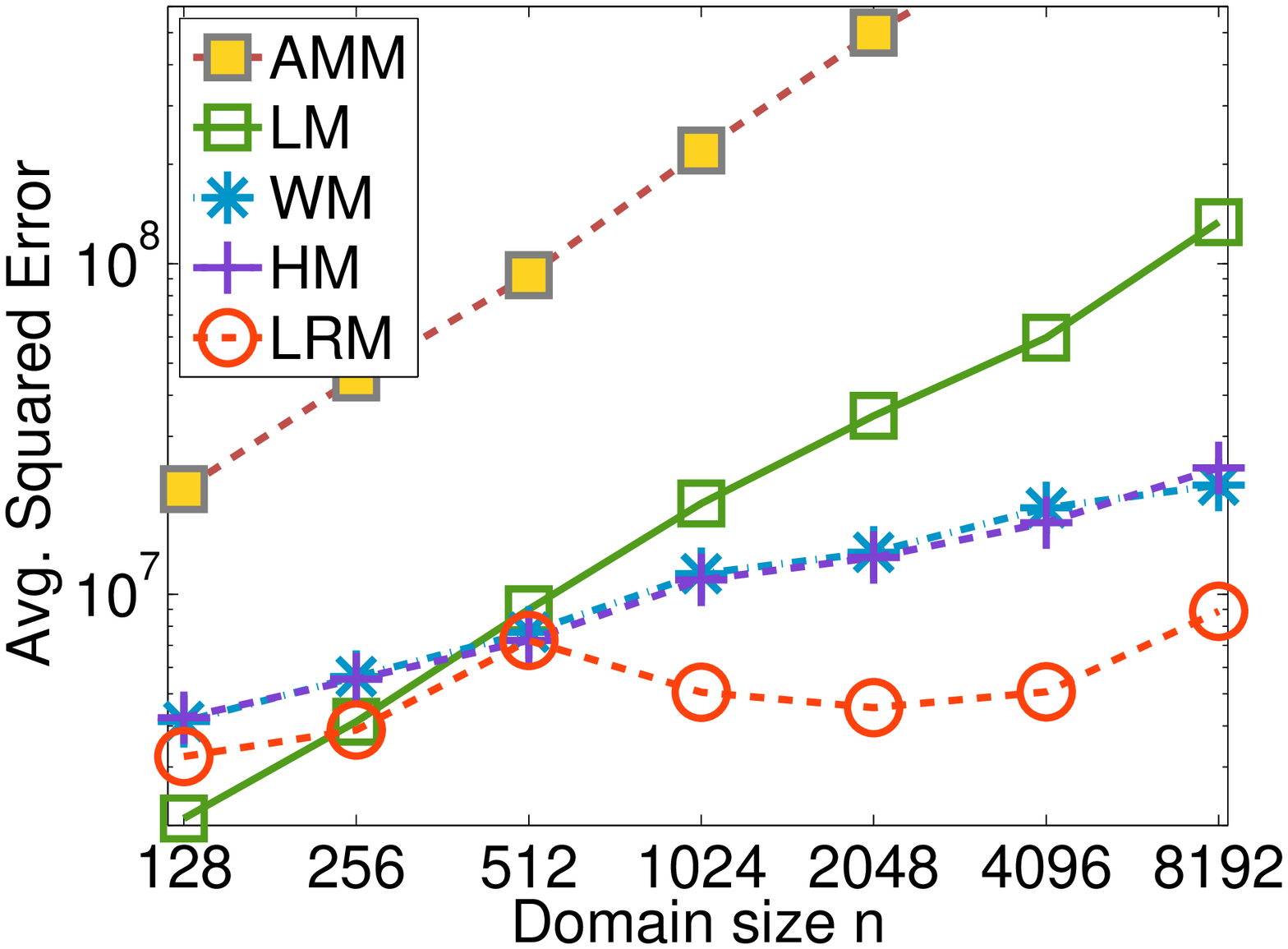}}
\vspace{-5pt}
\caption{Effect of domain size $n$ on workload \emph{WRange} with $\epsilon=0.1$}\label{fig:exp:n:WRange}
\vspace{-10pt}
\end{figure*}

\begin{figure*}[t]
\centering \subfigure[\emph{Search Logs}]
{\includegraphics[width=0.3\textwidth]{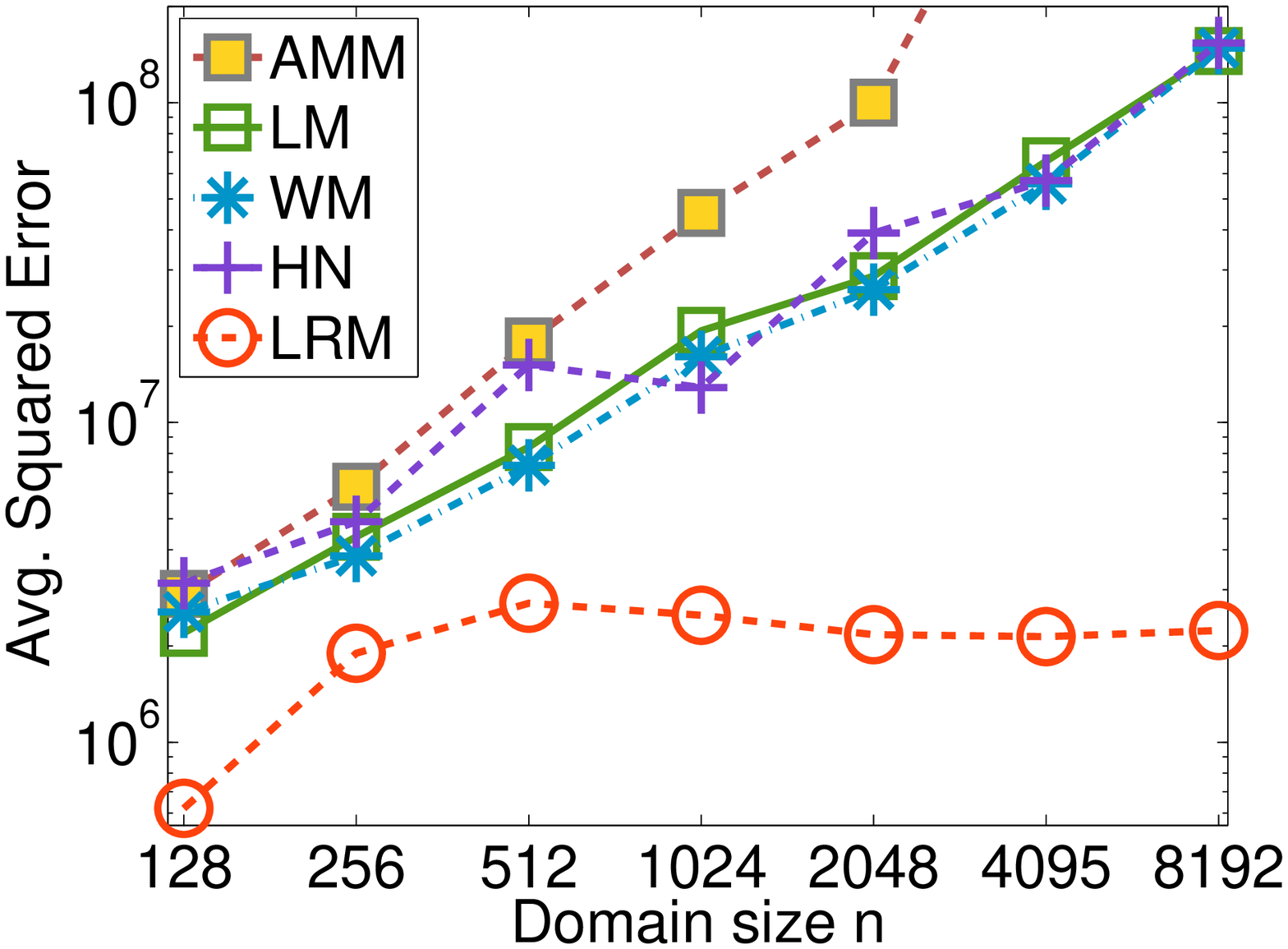}}
\subfigure[\emph{NetTrace}]
{\includegraphics[width=0.3\textwidth]{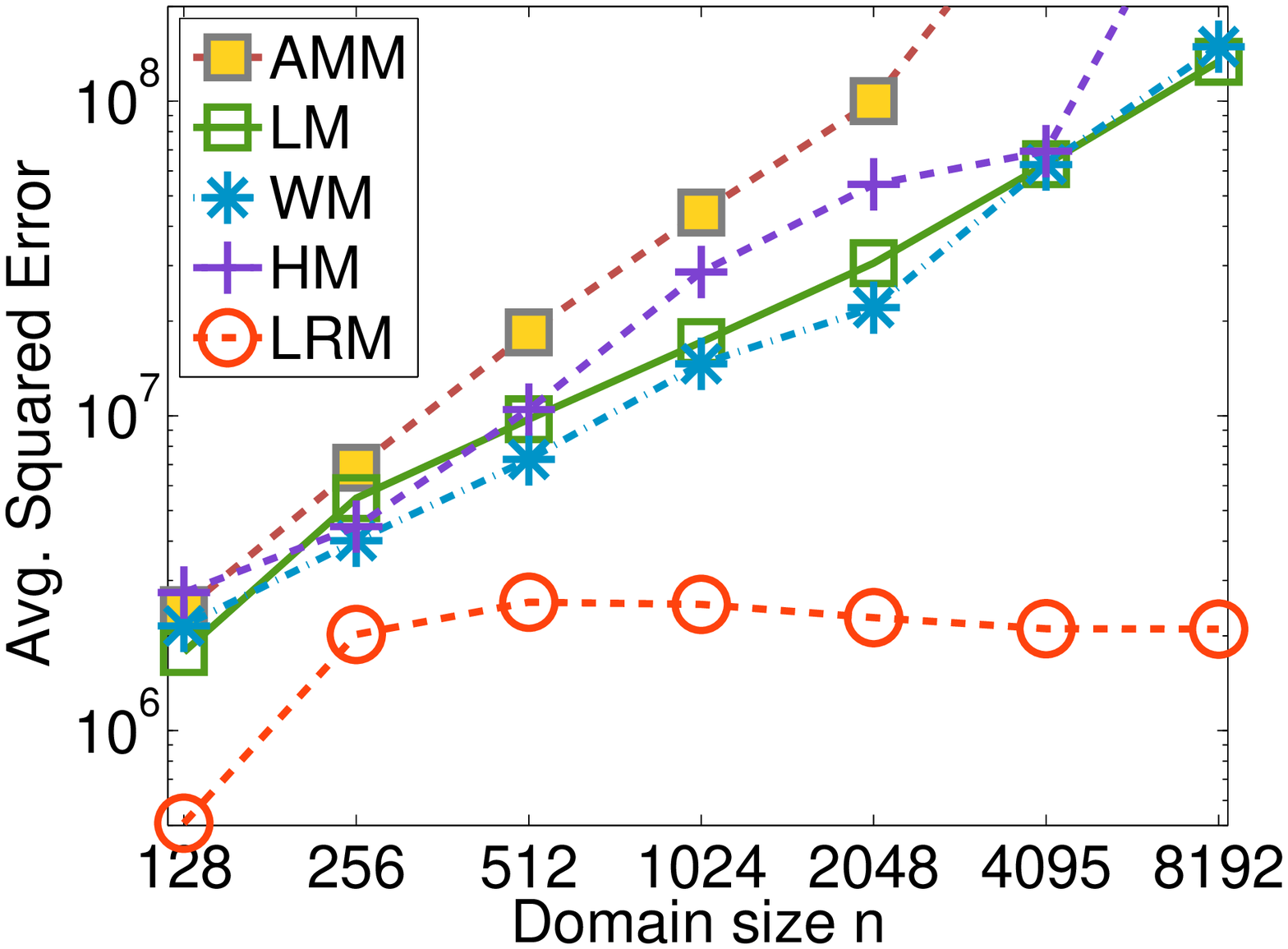}}
\centering \subfigure[\emph{Social Network}]
{\includegraphics[width=0.3\textwidth]{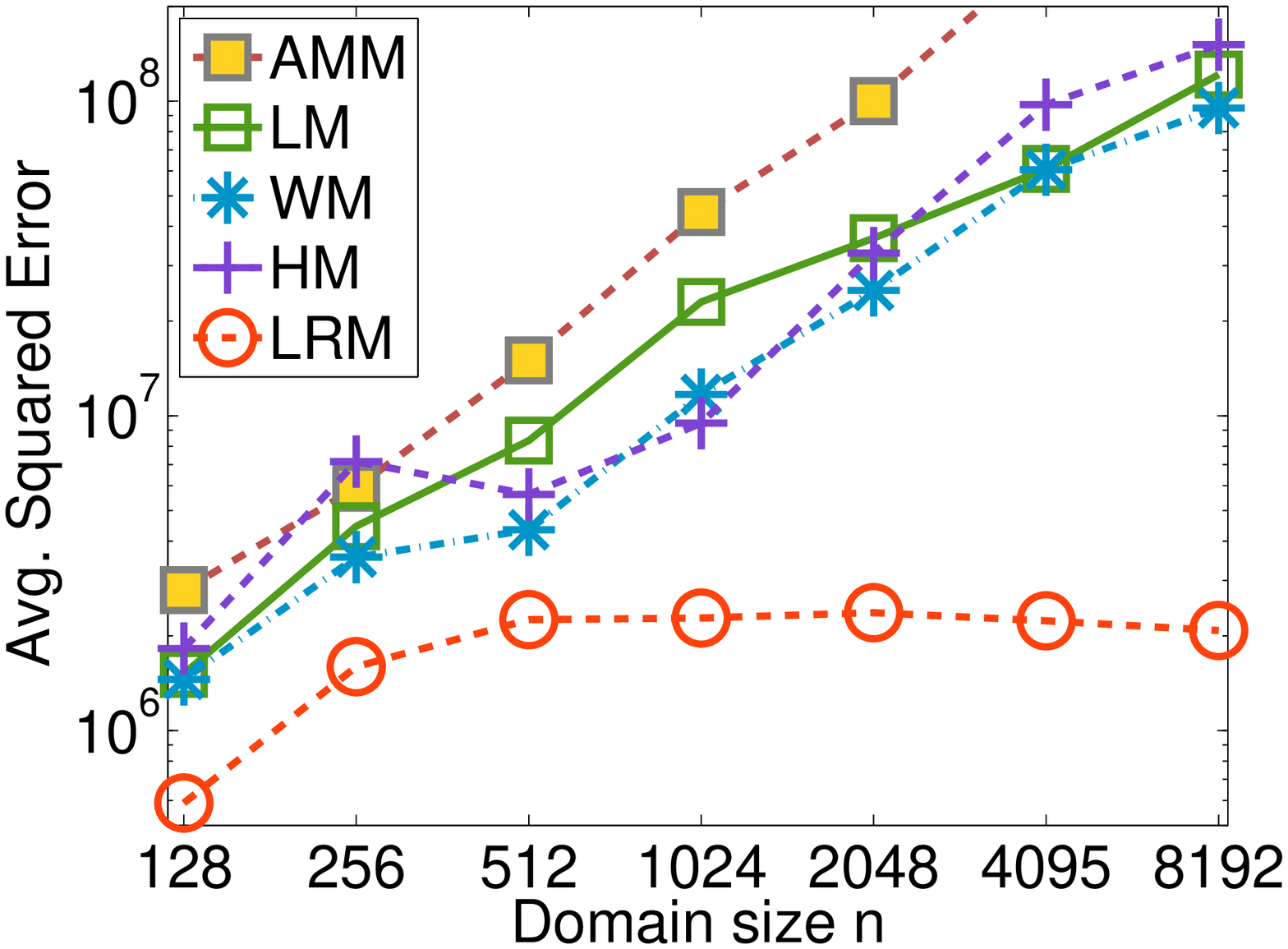}}
\vspace{-5pt}
\caption{Effect of domain size $n$ on workload \emph{WRelated} with $\epsilon=0.1$}\label{fig:exp:n:WRelated}
\vspace{-10pt}
\end{figure*}


\begin{figure*}[t]
\centering \subfigure[\emph{Search Logs}]
{\includegraphics[width=0.3\textwidth]{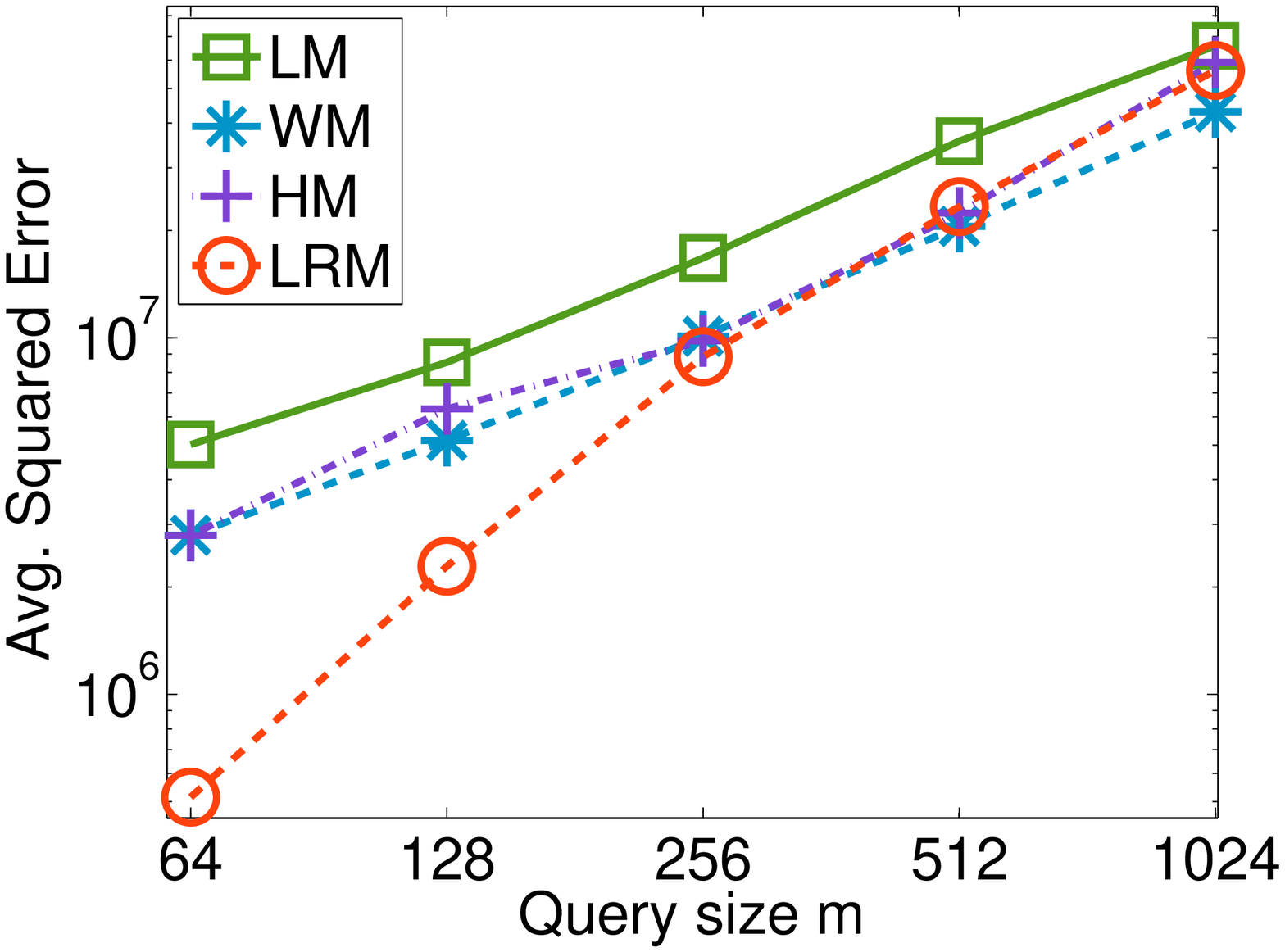}}
\subfigure[\emph{NetTrace}]
{\includegraphics[width=0.3\textwidth]{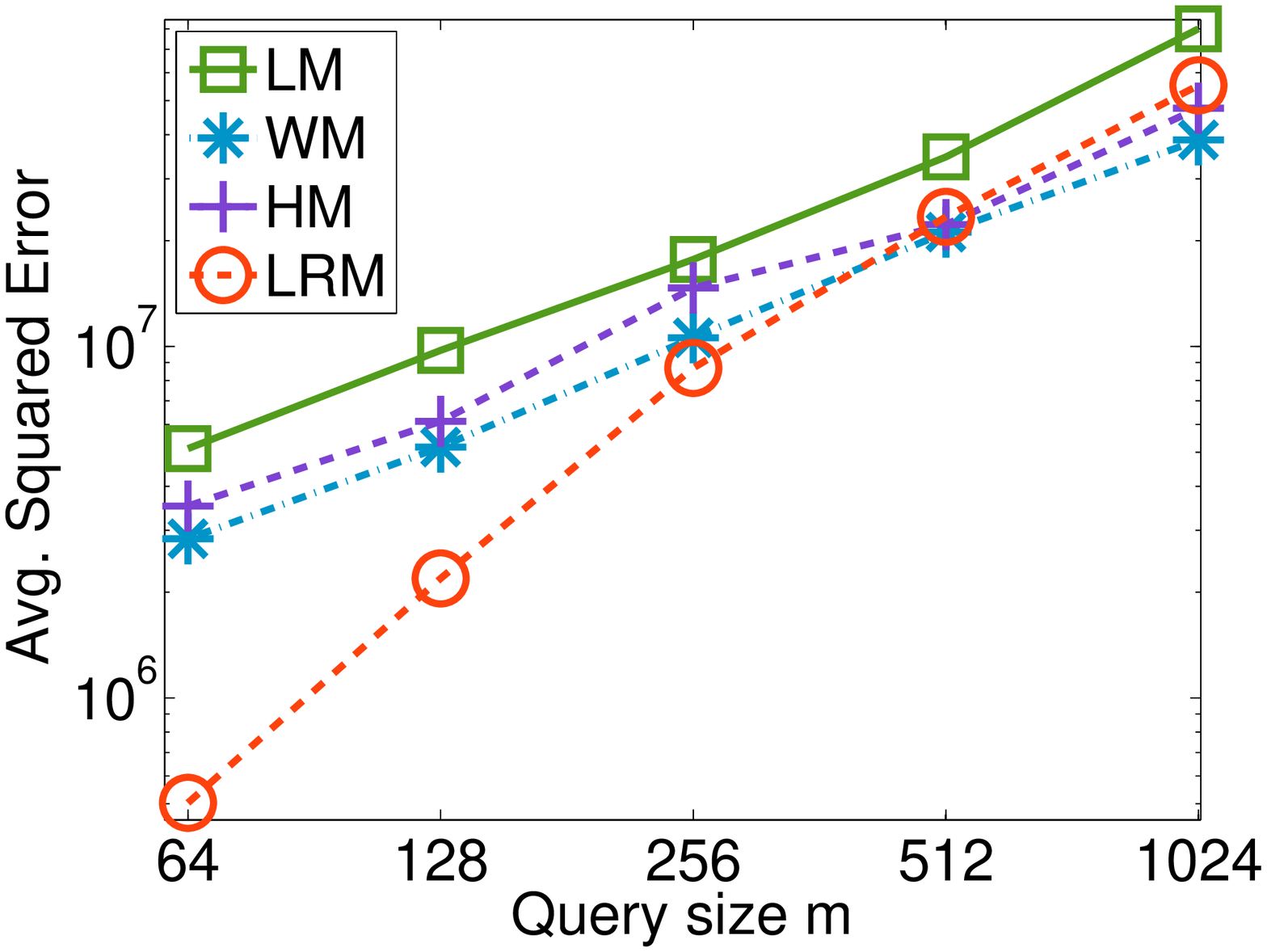}}
\centering \subfigure[\emph{Social Network}]
{\includegraphics[width=0.3\textwidth]{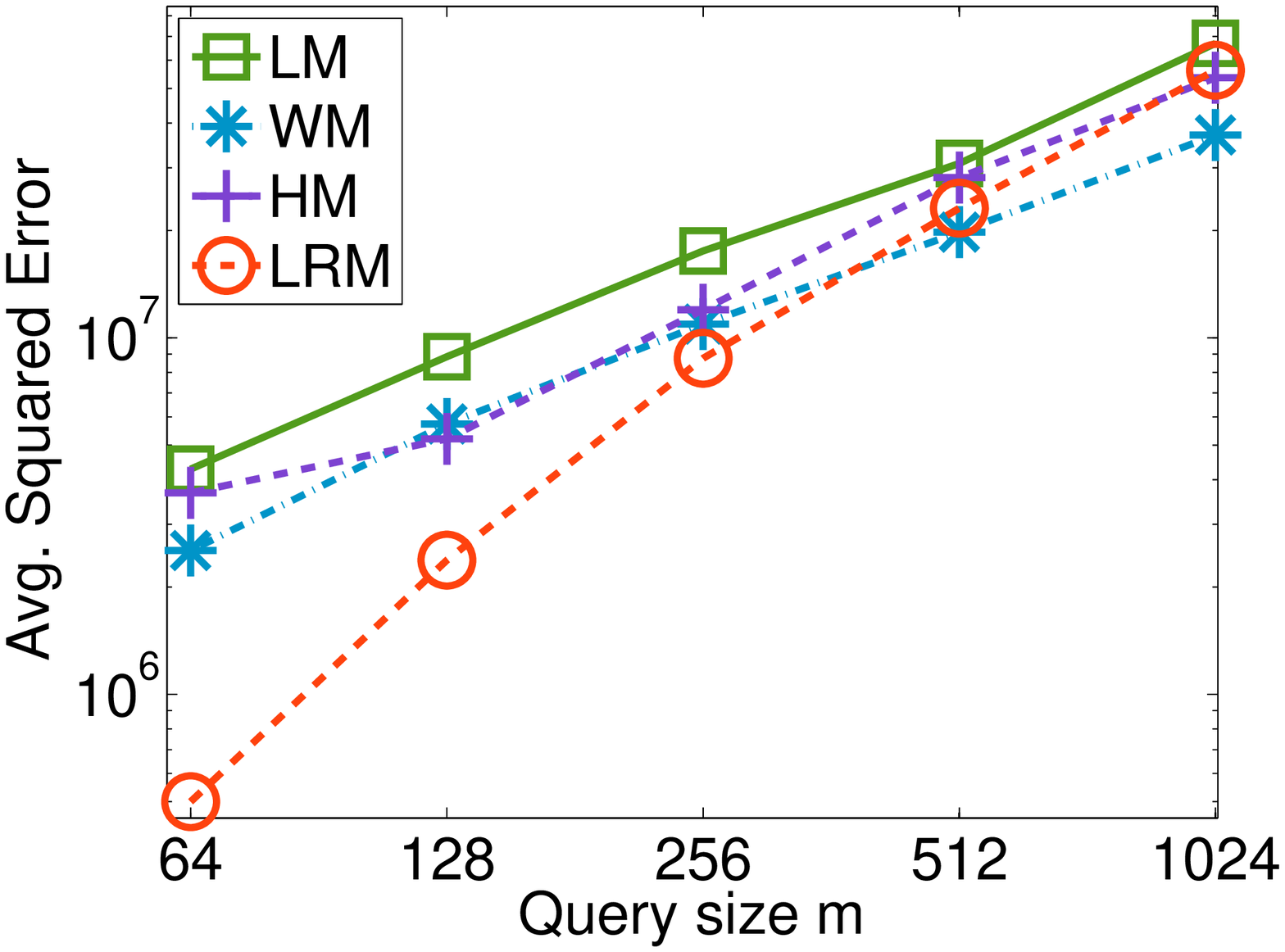}}
\caption{Effect of number of queries $m$ on workload \emph{WRange} with $\epsilon=0.1$}
\vspace{-5pt}
\label{fig:exp:m:WRange}
\vspace{-10pt}
\end{figure*}

\begin{figure*}[t]
\centering \subfigure[\emph{Search Logs}]
{\includegraphics[width=0.3\textwidth]{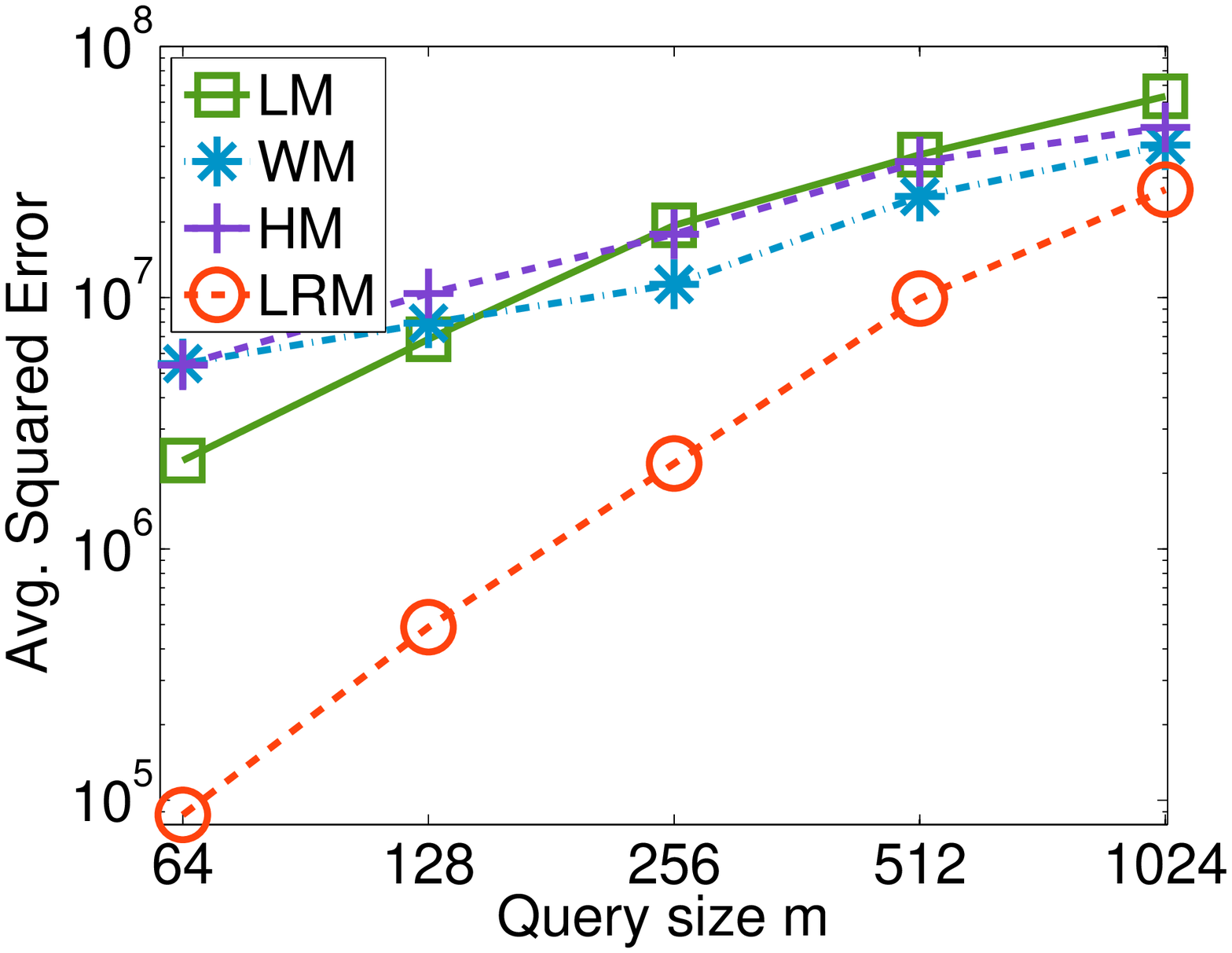}}
\subfigure[\emph{NetTrace}]
{\includegraphics[width=0.3\textwidth]{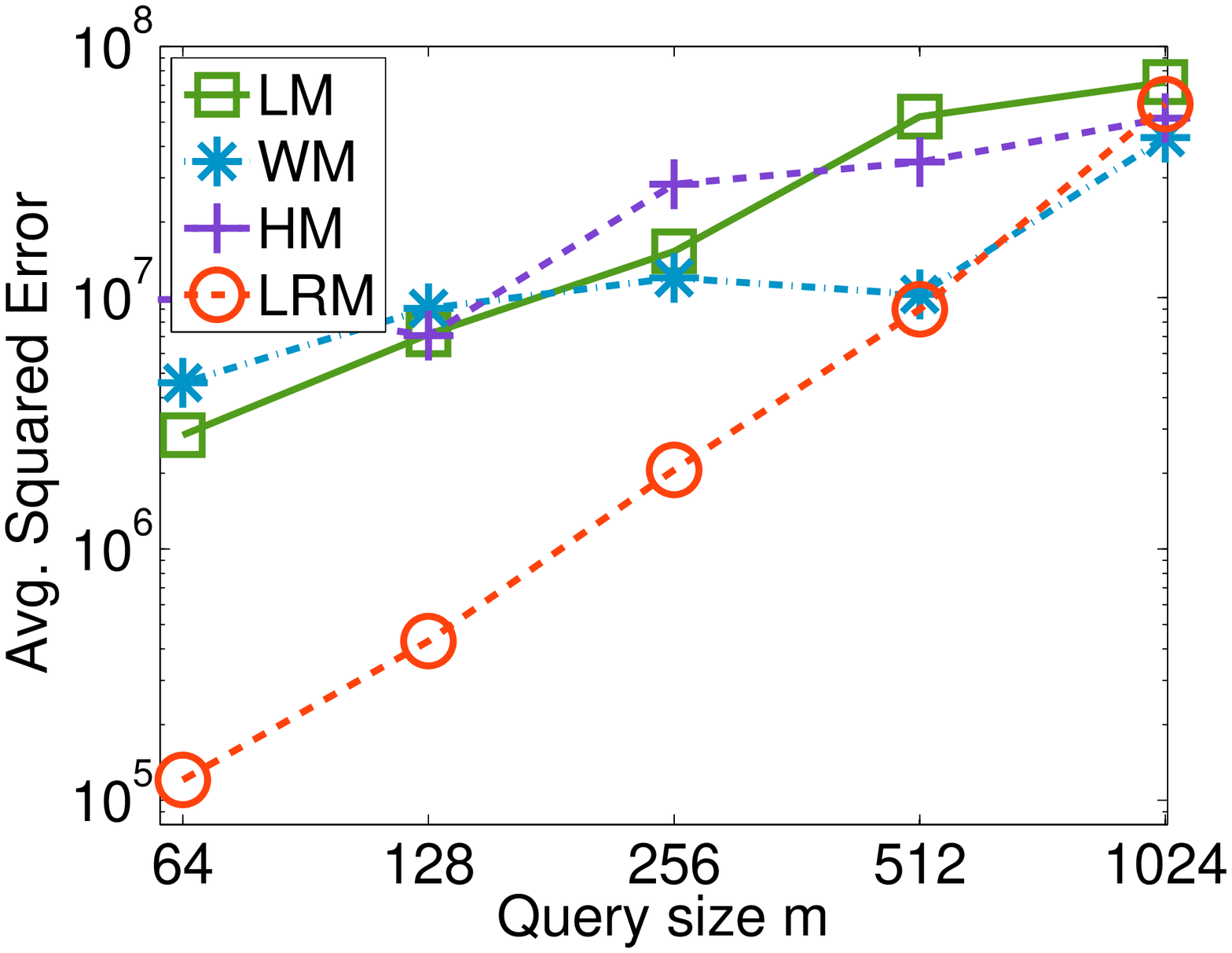}}
\centering \subfigure[\emph{Social Network}]
{\includegraphics[width=0.3\textwidth]{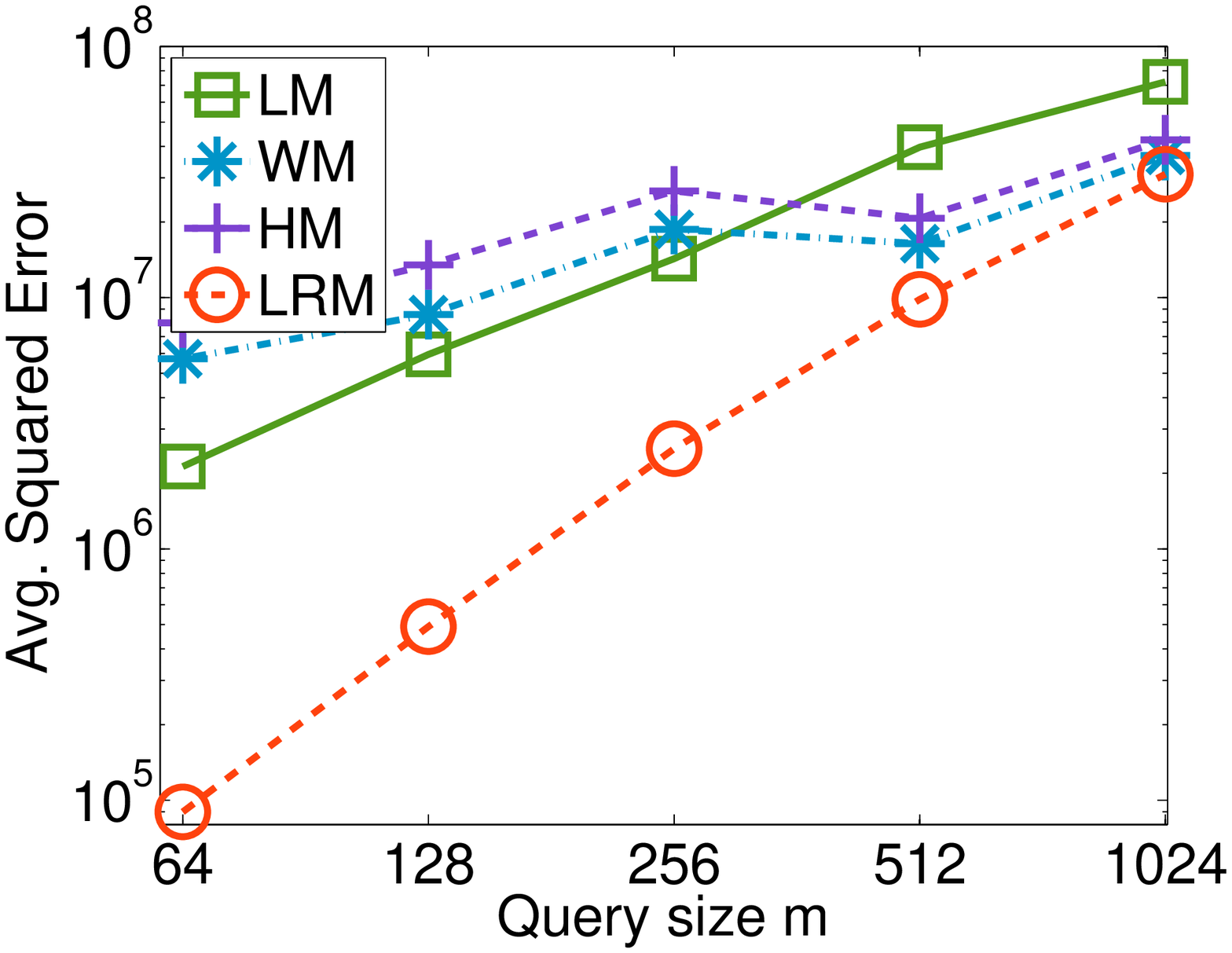}}
\vspace{-5pt}
\caption{Effect of number of queries $m$ on workload \emph{WRelated} with $\epsilon=0.1$}
\label{fig:exp:m:WRelated}
\vspace{-10pt}
\end{figure*}

\begin{figure*}[t]
\centering \subfigure[\emph{Search Logs}]
{\includegraphics[width=0.3\textwidth]{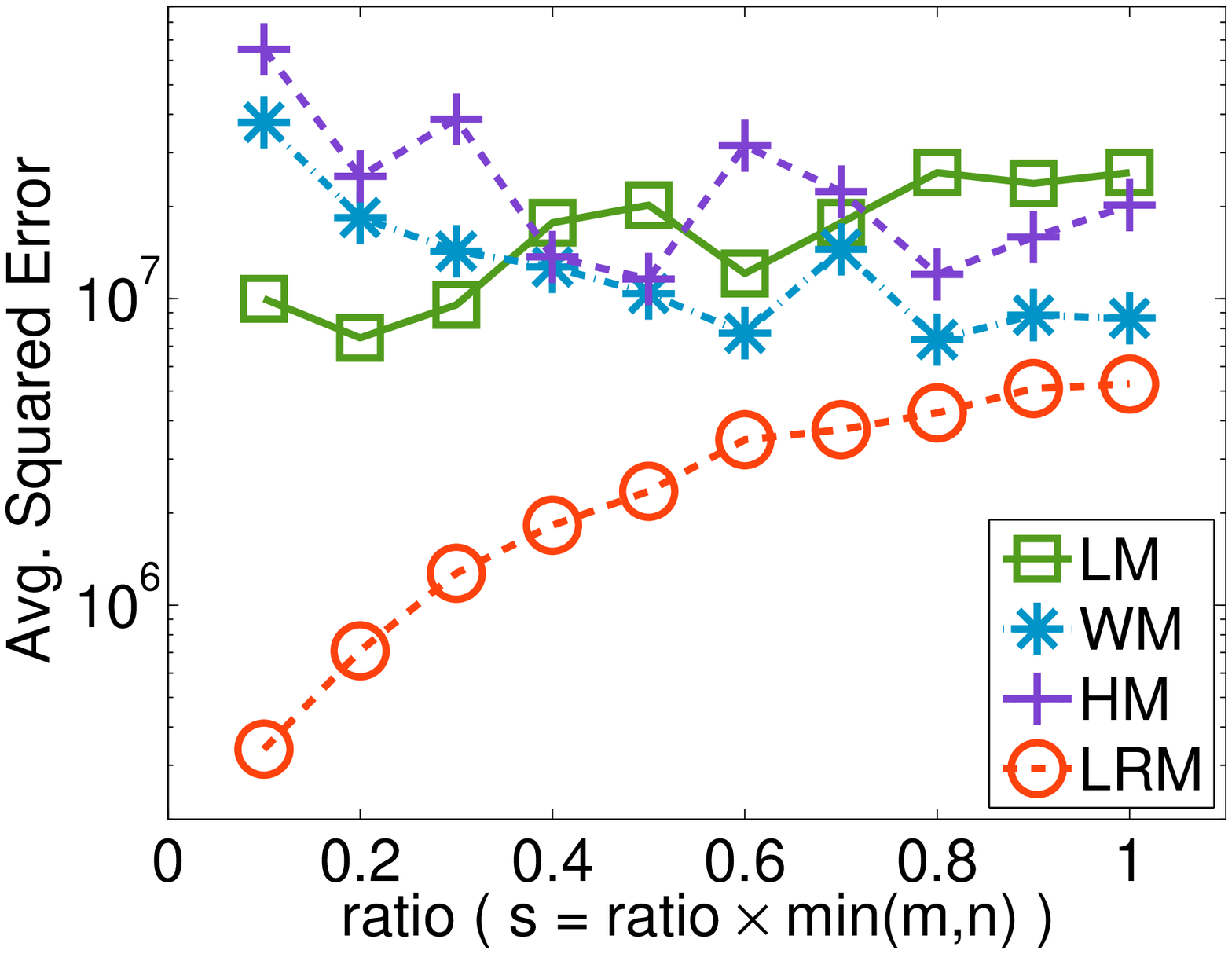}}
\subfigure[\emph{NetTrace}]
{\includegraphics[width=0.3\textwidth]{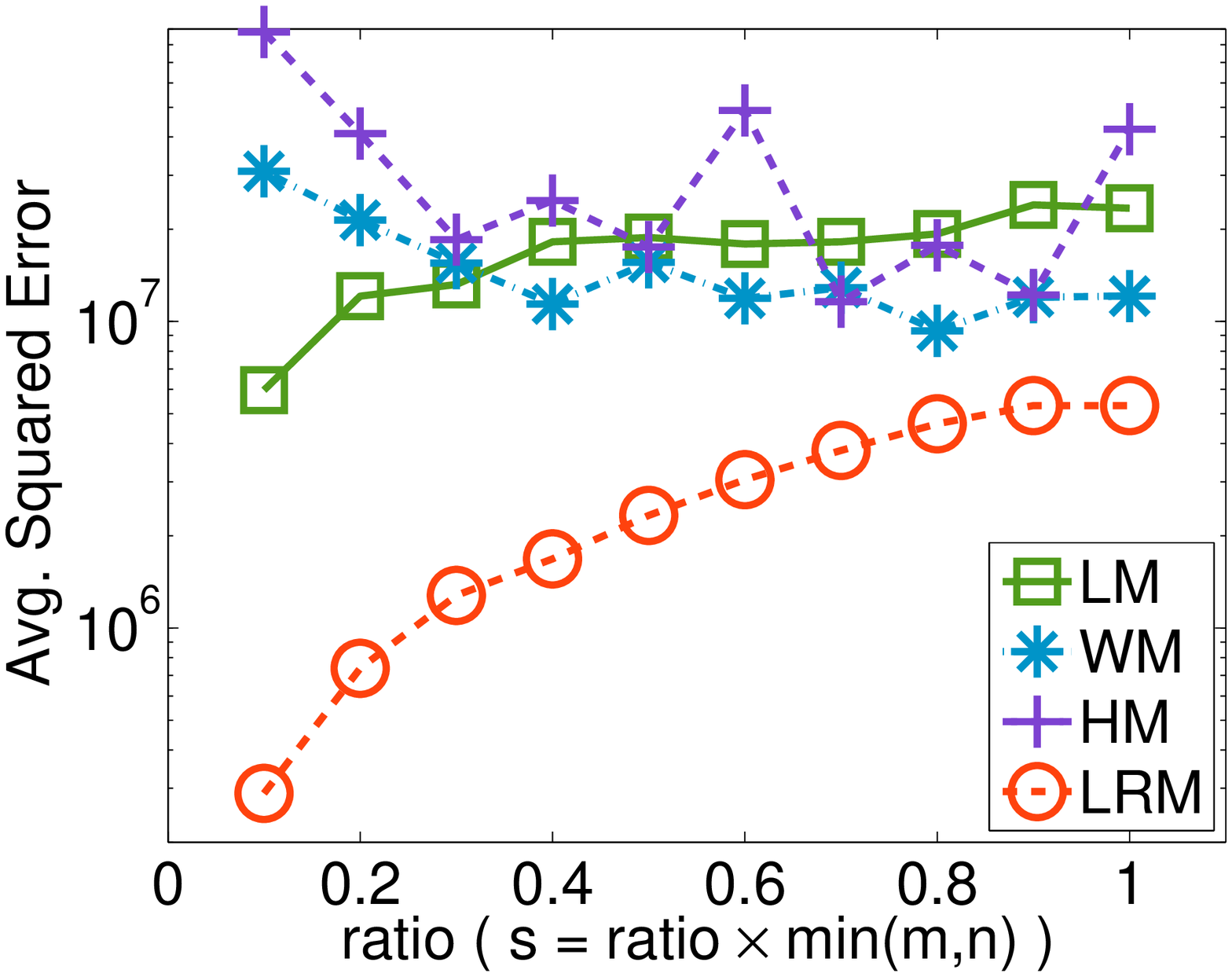}}
\centering \subfigure[\emph{Social Network}]
{\includegraphics[width=0.3\textwidth]{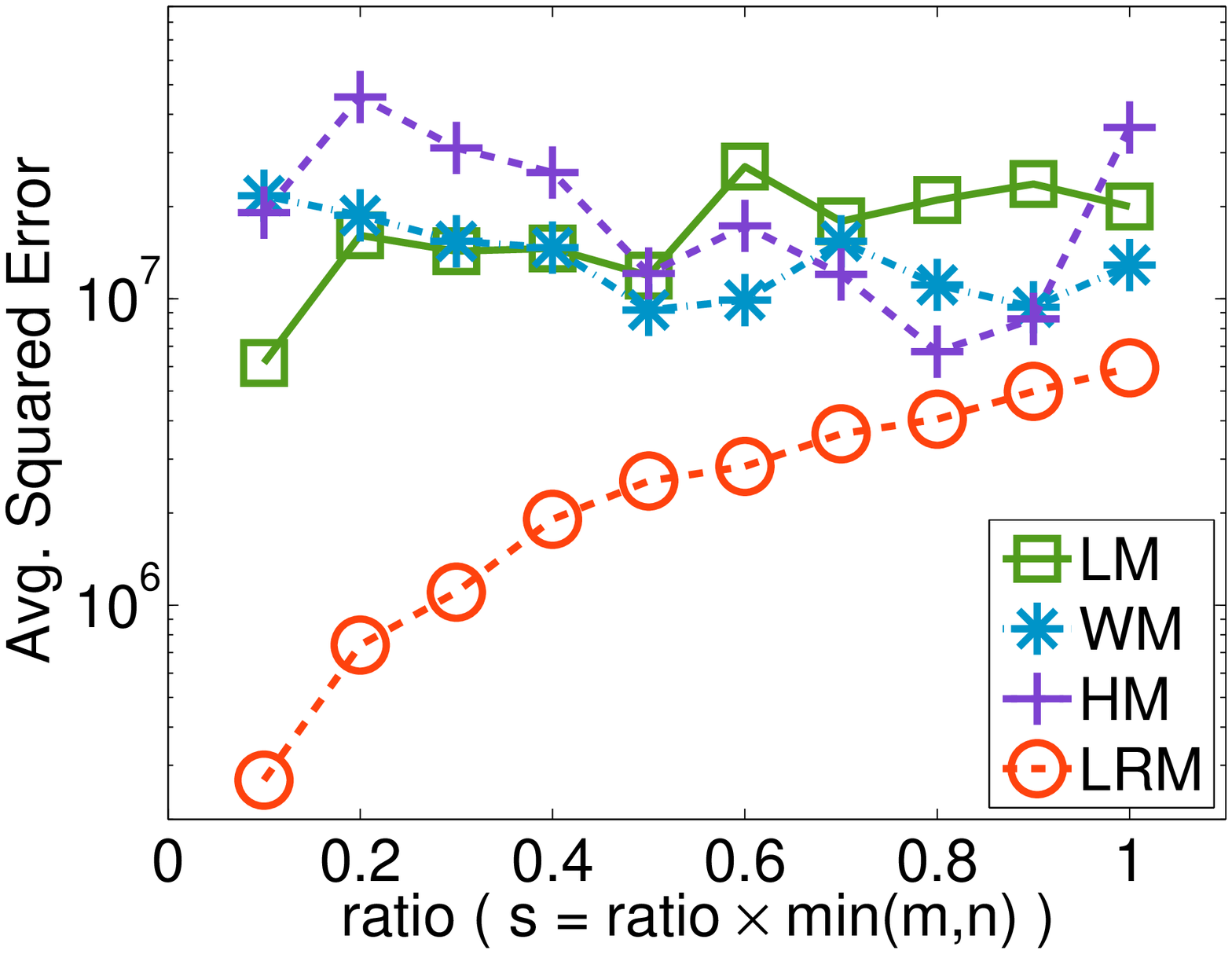}}
\vspace{-5pt}
\caption{Effect of parameter $s$ with $\epsilon=0.1$}
\label{fig:exp:s}
\vspace{-10pt}
\end{figure*}


\section{Experiments}\label{sec:exp}

This section demonstrates the effectiveness of the proposed Low-Rank Mechanism (LRM), and compares it against four state-of-the-art
methods: {the approximate Matrix Mechanism (AMM) that optimizes the $\mathcal{L}_2$
approximation \cite{LHR+10}}, the Laplace Mechanism (LM) \cite{DMNS06}, the Wavelet Mechanism (WM) \cite{XWG10} and the Hierarchical Mechanism (HM) \cite{HRMS10}. 
The details of our AMM implementation are available in Appendix \ref{sec:appedinex:matrix}. All methods were implemented and tested in Matlab on a desktop PC with Intel quad-core 2.50 GHz CPU and 4GBytes RAM. In all experiments, every algorithm is executed
20 times and the average performance is reported. We employ three
popular real datasets used in \cite{HRMS10,XZXYY11}:
\emph{Search Log}, \emph{Net Trace} and \emph{Social Network}.
\emph{Search Log} includes search keyword statistics
collected from \emph{Google Trends} and \emph{American Online}
between 2004 and 2010. \emph{Social Network} gives the number of
users in a social network site with specific degrees in the social
graph. \emph{Net Trace} is a statistical database containing the number of
TCP packets related to particular IP addresses, which is collected from a university intranet. \emph{Search Logs}, \emph{Net
Trace} and \emph{Social Network} contain $2^{16} = 65,536$, $2^{15} =
32,768$ and $11,342$ entries respectively. The reader is referred to \cite{HRMS10} for more details of these datasets. We published our Matlab implementations of all algorithms used in the experiments, as well as sample datasets, online at \url{http://yuanganzhao.weebly.com/}.

To evaluate the impact of data domain cardinality on real datasets, we transform the original counts into a vector
of fixed size $n$ (domain size), by merging consecutive
counts in order. Given the number $m$ of linear queries in the batch, we generate three different
types of workloads, namely \emph{WDiscrete}, \emph{WRange} and
\emph{WRelated}. In \emph{WDiscrete}, for each weight $W_{ij}$ of
query $q_i$ in the batch, we randomly select $W_{ij}=1$ with
probability 0.02 and set $W_{ij}=-1$ otherwise. In \emph{WRange}, a
batch of range queries on the domain are generated, by randomly
picking up the starting location $a$ and ending location $b$
following a uniform distribution on the domain. Given the interval
$(a,b)$, we set $W_{ij}$ of query $q_i$ in the batch to 1 for every $a\leq j\leq b$ and all other
weights to 0. Finally, for \emph{WRelated}, we generate $s$ (discussed later) independent base queries $A$ of size $s\times n$,
by randomly assigning weights to the queries under a standard $(0,1)$-normal distribution. Another group of correlation
matrix $C$ of size $m\times s$ are generated similarly. The final
workload $W$ of size $m\times n$ is the product of $C$ and $A$.


We test the impact of five parameters in our experiments:
$\gamma$, $r$, $n$, $m$ and $s$. $\gamma$ is the relaxation factor
defined in Formula (\ref{eqn:relaxed-problem}). $r$ is the
number of columns in $B$ (and also the number of rows in $L$). $n$
is the size of the domain and $m$ is the number of queries in the
batch. Finally, $s$ is the number of rows of queries in the base
$A$, which is only used in the generation of \emph{WRelated}. The range of
all these five parameters is summarized in Table
\ref{tab:exp:parameters}. Unless otherwise specified, the default
parameters in bold are used. Moreover, we test three different
privacy budgets, $\epsilon=1$, $0.1$ and $0.01$. Note that the
squared error incurred by all the methods is quadratic in
$1/\epsilon$.

\begin{table}[t]
\small
\begin{center}
\begin{tabular}{|c|c|}
\hline
$\gamma $ & $  0.0001, 0.001, \textbf{0.01}, 0.1, 1, 10$\\
\hline
$ r $ & $\{0.8, 1.0, \textbf{1.2}, 1.4, 1.7, 2.1, 2.5, 3.0, 3.6\} \times rank(W) $\\
\hline
$ n $ & $ 128, 256, 512, \textbf{1024}, 2048, 4096, 8192$\\
\hline
$ m$ & $64, 128, \textbf{256}, 512, 1024$\\
\hline
$ s $ & $ \{0.1, 0.2, 0.3, 0.4, \textbf{0.5}, 0.6, 0.7, 0.8, 0.9, 1.0\} \times min(m,n)$\\
\hline
\end{tabular}
\vspace{-5pt} \caption{Parameters used in the experiments}
\label{tab:exp:parameters} \vspace{-10pt}
\end{center}
\end{table}

In the experiments, we measure \emph{Average Squared Error} and
\emph{Computation Time} of the methods. Specifically, the
\emph{Average Squared Error} is the average squared $\mathcal{L}_2$
distance between the exact query answers and the noisy answers. In
the following, we first examine the impact of $\gamma$ and $r$,
which are only used in the LRM method. The results provide important insights on how
to tune these two parameters to maximize the utility of the LRM
method.

\subsection{Impact of $\gamma$ and $r$ on LRM}\label{vary_gamma}

In LRM, $\gamma$ is an important parameter controlling the
relaxation on the approximation of $BL$ to $W$. In our first set of
experiments, we investigate the impact of $\gamma$ on the accuracy
and the efficiency of LRM. Figure \ref{fig:exp:gamma} reports
the performance of LRM under all three different workloads,
\emph{WDiscrete}, \emph{WRange} and \emph{WRelated} on the \emph{Search
Log} dataset with varying values for $\gamma$. The results in the figure show that
the errors of LRM on all three workloads are not sensitive to $\gamma$ in the range from $10^{-4}$ to $10$. On the other hand, LRM executes much faster with larger $\gamma$. This suggests that a larger value for $\gamma$ is preferred in practice, to achieve high efficiency without losing much on result accuracy. Moreover, we also test with three different values of the privacy budget $\epsilon$. Since the decomposition method does not
rely on $\epsilon$, the shapes of the result curves with different $\epsilon$ values are nearly identical, albeit at different scales. The average error is quadratic in the privacy budget
$\frac{1}{\epsilon}$, as expected.

In LRM, $r$ is another important parameter that determines the rank of
the matrix $BL$ that approximates the workload $W$. $r$ affects both the
approximation accuracy and the optimization speed. When $r$ is too small, e.g., when $r<rank(W)$, our optimization formulation may fail to find a good approximation, leading to suboptimal
accuracy for the query batch. On the other hand, an overly large $r$ leads to poor efficiency, as the search space expands dramatically. We thus test LRM
with varying $r$, by controlling the ratio of $r$ to the actual rank
$rank(W)$, on the \emph{Search Log} dataset. We record the average
squared error under all the workloads and report it in Figure
\ref{fig:exp:r}.

There are several important observations in Figure \ref{fig:exp:r}.
First, when $r<rank(W)$, the accuracy of LRM is far worse (up to two orders of magnitude) than that in other settings. Second, the performance of LRM is rather stable
when $r$ becomes larger than $1.2\cdot rank(W)$. This
is because the optimization formulation has enough freedom to find
the optimal decomposition when $r$ is larger than $rank(W)$.
Finally, the amount of computation spent on workload decomposition increases
exponentially with $r$. Thus, to balance the efficiency and
effectiveness of LRM, a good value for $r$ is between $rank(W)$
and $1.2\cdot rank(W)$. We use the latter as the default value in the subsequent experiments.

\subsection{Impact of Varying Domain Size $n$}

We now evaluate the performance of all mechanisms with varying domain size $n$. As mentioned earlier in this section, the domain size is controlled by merging
consecutive counts in the original domain. While different
workloads and datasets are used, we only test settings with $\epsilon=0.1$
because $\epsilon$ does not have much impact on the relative performance of different mechanisms. In
Figures \ref{fig:exp:n:WDiscrete}, \ref{fig:exp:n:WRange} and
\ref{fig:exp:n:WRelated}, we report the result error rates of all these
mechanisms.

In all experiments, the approximate Matrix Mechanism (AMM) is much worse than the other mechanisms, sometimes by an order of magnitude. This is mainly because 
the $\mathcal{L}_2$ approximation used by AMM does not lead to a good optimization of the actual objective function formulated using the error measure in $\mathcal{L}_1$. Because of its poor performance, we exclude AMM in the rest of the
experiments.

On the \emph{WDiscrete} workload, the Laplace Mechanism (LM) outperforms all other mechanisms when the
domain size is relatively small. This is in part due to the fact that the Wavelet Mechanism (WM) and the Hierarchical Mechanism (HM) are mainly designed
to optimize range queries. While all other mechanisms incur linear
error in terms of the domain size $n$, LRM's error stops increasing
when the domain size is larger than 512. This is because
LRM's error relies on the rank of the workload matrix $W$, and
$rank(W)$ is no larger than $\min(m,n)$ no matter how large $n$ is.
This explains the excellent performance of LRM on larger domains. On
the \emph{WRange} workload, the errors of WM and HM are smaller than LM
when the domain size is no smaller than 512, in which case their
strategies work better. LRM's performance is still
significantly better than any of them, since LRM fully utilizes the
correlations between these range queries on large domains. Finally, on the \emph{WRelated} workload, LRM achieves the
best performance on all test cases. The performance gap between LRM and other methods is over two orders of magnitude, when the domain size reaches
8192. Since \emph{WRelated} naturally leads to a low rank workload matrix $W$, this result verifies LRM's vast benefit from exploiting the low-rank property of the workload.

\subsection{Impact of Varying Query Size $m$}

In this subsection, we test the impact of the query set cardinality
$m$ on the performance of the mechanisms. We mainly focus on
settings when the number of queries $m$ is no larger than the domain size
$n$, i.e. $m \leq n$. Due to space limitations, we only
present the results on \emph{WRange} and \emph{WRelated} workloads
in Figures \ref{fig:exp:m:WRange} and \ref{fig:exp:m:WRelated}.

The results lead to several interesting observations. On
\emph{WRange} workload (Figure \ref{fig:exp:m:WRange}), LRM
outperforms the other mechanisms, when the number of queries $m$ is
significantly smaller than $n$. With growing $m$, the performance of
all mechanisms on \emph{WRange} tends to converge. When $m=1024$, WM
achieves the best performance among all mechanisms, since it is
optimized for range queries. The degeneration in performance of LRM
is due to the lack of low rank property when the batch contains too
many random range queries. On \emph{WRelated} workload, LRM is
dramatically better than the other methods, for any query set
cardinality $m$. Regardless of the value of $m$, the rank of the
\emph{WRelated} workload $W$ remains low, which is solely determined
by the parameter $s$ used in the workload generation procedure.
These results further confirm that the squared error generated by
LRM scales linearly with the rank of the workload.

\subsection{Impact of Varying Query Rank $s$}

All previous experiments demonstrate LRM's substantial performance
advantage when the workload matrix has low rank. In this group of
experiments, we manually control the rank of workload $W$ to verify
the correctness of our claim. Recall that the parameter $s$ determines
the size of the matrix $C_{m\times s}$ and the size of the matrix
$A_{s\times n}$ in the generation of the \emph{WRelated} workload. When
$C$ and $A$ contain only independent rows/columns, $s$ is exactly
the rank of the workload matrix $W=CA$. In Figure \ref{fig:exp:s},
we vary $s$ from $0.1\min(m,n)$ to $\min(m,n)$. Compared to the other mechanisms, LRM maintains an accuracy advantage of over two orders of magnitude, when
the rank of the workload matrix is low. With increasing
rank of $W$, the accuracy of other mechanisms remain stable, while LRM's error grows rapidly. This phenomenon again confirms that the low rank property is the main reason behind LRM's advantages with respect to error minimization.

\section{Conclusion}\label{sec:concl}

This paper presented the \emph{Low Rank Mechanism} (LRM), an optimization framework that minimizes the overall error in the results of a batch of linear queries under $\epsilon$-differential privacy. LRM
is the first practical method for a large number of linear
queries, with an efficient and effective implementation using well
established optimization techniques. Experiments show that LRM
significantly outperforms other state-of-the-art differentially
private query processing mechanisms, often by orders of magnitude.
The current design of LRM focuses on exploiting the correlations
between different queries. One interesting direction for future work
is to further optimize LRM by utilizing also the correlations
between data values, e.g., as is done in \cite{XZXYY11, RN10, LZMY11}.

\section*{Acknowledgments} Yuan and Hao are supported by NSF-China (61070033,
61100148), NSF-Guangdong (9251009001000005, S2011040004804) and Key Technology Research and Development Programs of Guangdong Province (2010B050400011). Zhang, Winslett, Xiao and Yang are supported by SERC 102-158-0074 from Singapore's A*STAR.
Xiao is also supported by SUG Grant M58020016 and AcRF Tier 1 Grant RG 35/09 from Nanyang Technological University.

\bibliographystyle{abbrv}

\normalsize

\appendix
\section{Proofs}

\noindent\textit{Lemma \ref{lem:decomp_error}: }
\begin{proof}
Based on the definition of the mechanism in Eq.
(\ref{eqn:part_mech}), the residual of the noisy result with respect
to the exact result, i.e. $Q(D)-M_P(Q,D)$, is $B\cdot
Lap\left(\frac{\Delta(B,L)}{\epsilon}\right)^r$. The expected
squared error is thus $\sum_{ij}B^2_{ij}\frac{2(\Delta(B,L))^2}{\epsilon^2}$. Since
$\Phi(B,L)=\sum_{ij}B^2_{ij}$, the expected error of the mechanism
is $2\phi(B,L)(\Delta(B,L))^2/\epsilon^2$.
\end{proof}

\noindent\textit{Lemma \ref{lem:rescale}:}
\begin{proof}
Based on the definition of sensitivity, we have
$\Delta(B',L')$\\$=\max_j\sum_i
|L'_{ij}|=\max_j\sum_i|L_{ij}/\alpha|=\alpha^{-1}\Delta(B,L)$.

The last equality holds because $\alpha$ is a positive constant. On the other hand, the scales of the decompositions follow a similar relationship:
$$\Phi(B',L')=\sum_{ij}(B'_{ij})^2=\sum_{ij}\alpha^2(B_{ij})^2=\alpha^2\Phi(B,L)$$

Therefore, $\Phi(B',L')(\Delta(B',L')^2=\Phi(B,L)(\Delta(B,L))^2$. Finally, since $B'L'=BL=W$, we reach the conclusion of the lemma.
\end{proof}

\noindent\textit{Theorem \ref{the:main}:}
\begin{proof}
Assume that $(B^*,L^*)$ is the best matrix decomposition for minimizing the expected squared error for $M_P(Q,D)$. In the following, we prove that $(B^*,L^*)$ is optimal, if and only if it also minimizes the program in Formula (\ref{eqn:opt-problem}).

(\textit{if} part): If $(B,L)$ minimizes Formula (\ref{eqn:opt-problem}) but $(B,L)$ incurs more expected error than $(B^*,L^*)$, implying that
$$\Phi(B^*,L^*)(\Delta(B^*,L^*))^2<\Phi(B,L)(\Delta(B,L))^2$$

By applying Lemma \ref{lem:rescale}, we can construct another
decomposition $B'=\Delta(B^*,L^*)B^*$ and
$L'=\Delta(B^*,L^*)^{-1}L^*$, such that
$\Phi(B',L')(\Delta(B',L'))^2<\Phi(B,L)(\Delta(B,L))^2$. On the
other hand, since $\Delta(B',L')\leq 1$, we have $\max_j\sum_i
|L'_{ij}|= 1$. Therefore, we can derive the following inequalities.
\begin{eqnarray}
\Phi(B',L') &=& \Phi(B',L')(\Delta(B',L'))^2\nonumber\\
  &<& \Phi(B,L)(\Delta(B,L))^2\nonumber\\
  &\leq& \Phi(B,L)\nonumber
\end{eqnarray}
Finally, since $\Phi(B',L')=\mbox{tr}(B'^TB')$ and $\Phi(B,L)=\mbox{tr}(B^TB)$, it leads to a contradiction if $\mbox{tr}(B'^TB')<\mbox{tr}(B^TB)$.

(\textit{only if} part): If $(B^*,L^*)$ is not the optimal solution to the program in Formula (\ref{eqn:opt-problem}), the optimal solution $(B,L)$ must incur less expected error, using a similar strategy. This completes the proof of the theorem.
\end{proof}

\noindent\textit{Lemma \ref{lem:upper}:}
\begin{proof}
To prove the lemma, we aim to artificially construct a workload decomposition $W=BL$ satisfying the constraints of the optimization formulation. If the error of this artificial decomposition is no larger than the upper bound, the exact optimal solution must render results with less error.

Recall that $W$ has a unique SVD decomposition $W=U\Sigma V$ such
that $\Sigma$ is a diagonal matrix of size $r \times r$. We thus
build a decomposition $B=\sqrt{r}U\Sigma$ and
$L=\frac{1}{\sqrt{r}}V$, in which $r$ is the rank of the matrix $W$.
First, we will show such $(B,L)$ satisfies the constraints in
Formula (\ref{eqn:opt-problem}). It is straightforward to show it
satisfies the first constraint: $BL=\sqrt{r}U\Sigma
\frac{1}{\sqrt{r}}V=U\Sigma V=W$.

Regarding the second constraint, since $V$ only contains orthogonal
vectors, every column $j$ must have $\|V_{:j}\|_2 =\|v\|_2= 1$. By
the norm triangle inequality,
$\|v\|_2\leq\|v\|_1\leq\sqrt{r}\|v\|_2$, and we obtain
$\frac{1}{\sqrt{r}}\sum_i |V_{ij}|\leq 1$. Therefore, such $(B,L)$
must be a valid solution to the program.

The expected squared error of the artificial decomposition $W=BL$ is at most
\begin{eqnarray}
\mbox{tr}(B^TB)/\epsilon^{2}&=& \mbox{tr}((\sqrt{r}U\Sigma)^T(\sqrt{r}U\Sigma))/\epsilon^{2}\nonumber\\
 &=& \mbox{tr}(\Sigma^TU^TU\Sigma))r/\epsilon^2\nonumber\\
 &=& \sum^r_{k=1}\lambda^2_kr/\epsilon^2\nonumber
\end{eqnarray}

This proves that $\sum^r_{k=1}\lambda^2_kr/\epsilon^2$ is an upper
bound for the noise of our decomposition-based scheme.
\end{proof}

\noindent\textit{Lemma \ref{lem:lower}:}
\begin{proof}
In Corollary 3.4 in \cite{HT10}, Hardt and Talwar proved that any $\epsilon$-differential privacy mechanism incurs expected squared error no less than\footnote{\cite{HT10} used absolute error in the paper, which we change to squared error here.} $\Omega(r^3\left(Vol(PWB^n_1)\right)^{2/r}/\epsilon^2)$.

In the formula above, $B^n_1$ is the $\mathcal{L}_1$-unit ball.
$Vol(PWB^n_1)$ is the volume of the unit ball after the linear
transformation under $PW$, in which $P$ is any orthogonal linear
transformation matrix from $\mathbb{R}^m\mapsto\mathbb{R}^r$. To
prove the lemma, we construct an orthogonal transformation $P$ using
the SVD decomposition over $W=U\Sigma V$. By simply letting $P=U^T$,
since $U^TU$ and $VV^T$ are identity matrices,
we have $Vol(PWB^n_1)=Vol(PUVV^T\Sigma VB^n_1)=Vol(V(V^T\Sigma
V)B^n_1)=Vol(VB^n_1)\prod^r_{k=1}\lambda_k$. The last equality holds
due to Lemma 7.5 in \cite{HT10}. Consider the the convex body
$VB^n_1$. It is an $r$-dimensional unit ball after the orthogonal
transformation under $V$. Note that $Vol(B^r_1)$ can be computed using the well known $\Gamma$ function, as in \cite{wang2005}, $2^r\frac{\Gamma(2)}{\Gamma(1+r)}=\frac{2^r}{r!}$.
Therefore, the lower bound can be computed as: $
\Omega((\frac{2^r}{r!}\prod^r_{k=1}\lambda_k)^{2/r}r^3/\epsilon^2)
$. This reaches the conclusion of the lemma.
\end{proof}

\noindent\textit{Theorem \ref{the:opt}:}
\begin{proof}
To prove the theorem, we investigate the ratio of the upper bound to the lower bound.

\begin{eqnarray}
&& \frac{\sum^r_{k=1}\lambda^2_kr/\epsilon^2}{ \left(\frac{2^r}{r!}\prod^r_{k=1}\lambda_k\right)^{2/r}r^3/\epsilon^2 }\nonumber\\
&\leq&  \frac{\sum^r_{k=1}\lambda_1^2}{\left(\frac{2^r}{r!}\prod^r_{k=1}\lambda_r\right)^{2/r} r^2}\nonumber\\
&\leq& \frac{ r\lambda^2_1}{\left(\frac{2^r}{r!}\right)^{2/r}
\lambda_r^{2}r^2} =  \frac{C^2}{\left(\frac{2^r}{r!}\right)^{2/r}r}
\leq \left(\frac{C}{4}\right)^2r \nonumber
\end{eqnarray}

The last inequality holds due to the fact that $r!< \left(\frac{r}{2}\right)^r$ when $r>5$. Note that all the inequalities above are tight, and the equalities hold when $C=1$, i.e. $\lambda_1=\lambda_2=\ldots=\lambda_r$. Thus, we prove that the approximation factor of our decomposition scheme is $\mathcal{O}(C^2r)$.
\end{proof}

\noindent\textit{Theorem \ref{the:new_error}:}
\begin{proof}
When $W\neq BL$, the error has two parts. The first part is the
noises due to the Laplace random variables. Using Lemma
\ref{lem:decomp_error}, the incurred error is at most
$\frac{2}{\epsilon^2}\Phi(B,L)(\Delta(B,L))^2\leq
\frac{2}{\epsilon^2}\mbox{tr}(B^TB)$.

The second part of the error is the structural error on the results. The expected squared error is measured as
\begin{eqnarray}
& & ((W-BL)D)^T(W-BL)D\nonumber\\
&\leq& \|W-BL\|^2_FD^TD\nonumber =
\|W-BL\|^2_F\sum^n_{i=1}x^2_i\nonumber
\end{eqnarray}

The inequality is due to the Cauchy Schwartz inequality. By linearity of expectation, the expected squared errors can be simply summed up. This leads to the conclusion of the theorem.
\end{proof}
\noindent\textit{Theorem \ref{LowRankDPConvergence}: }
\begin{proof}
We use ${B^{(k)}}^*$ to denote the optimal solution of the Lagrangian
sub-problem in $k^{th}$ iteration. Note the following inequality on the sequence of the Lagrangian
subproblems:

\begin{eqnarray}\label{eqn:proof1}
& & \mathcal{J}({B^{(k+1)}}^*,{L^{(k+1)}}^*,{\pi^{(k)}}^*,\beta^{(k)}) \nonumber\\
&=& \min_{B,L} \mathcal{J}(B,L,{\pi^{(k)}}^*,\beta^{(k)}) \nonumber\\
&\leq& \min_{\|W-BL\|_F\leq\gamma, \forall j \sum_i |L_{ij}|\leq 1} \mathcal{J}(B,L,{\pi^{(k)}}^*,\beta^{(k)}) \nonumber \\
&=& \min_{\|W-BL\|_F\leq\gamma, \forall j \sum_i |L_{ij}|\leq 1}
\frac{1}{2}\mbox{tr}(B^TB) =\frac{1}{2}\mbox{tr}(B^{*T}B^*)\nonumber
\end{eqnarray}

Based on the above inequality, we derive the following inequality:
\begin{eqnarray}
& & \frac{1}{2}\mbox{tr}({B^{(k+1)}}^TB^{(k+1)})\nonumber\\
&=& \mathcal{J}({B^{(k+1)}}^*,{L^{(k+1)}}^*,{\pi^{(k)}}^*,\beta^{(k)}) - \langle \pi^{(k)}, W-B^{(k+1)}\nonumber\\
& & L^{(k+1)}\rangle + \frac{\beta^{(k)}}{2}\|W-B^{(k+1)}L^{(k+1)}\|_F^2\nonumber\\
&=& \mathcal{J}({B^{(k+1)}}^*,{L^{(k+1)}}^*,{\pi^{(k)}}^*,\beta^{(k)}) - \frac{1}{2\beta^{(k)}}(\|\pi^{(k)} + \beta^{(k)}\nonumber\\
& &   (W-B^{(k+1)}
L^{(k+1)})\|_F^2 - \|\pi^{(k)}\|_F^2 )\nonumber\\
&=& \mathcal{J}({B^{(k+1)}}^*,{L^{(k+1)}}^*,{\pi^{(k)}}^*,\beta^{(k)}) - \frac{1}{2\beta^{(k)}} (\|\pi^{(k+1)}\|_F^2 \nonumber\\
& &  - \|\pi^{(k)}\|_F^2) \nonumber\\
&\leq& \frac{1}{2}\mbox{tr}(B^{*T}B^*) - \frac{1}{2\beta^{(k)}}
\left(\|{\pi^{(k+1)}}^*\|_F^2 - \|{\pi^{(k)}}^*\|_F^2\right)
\nonumber
\end{eqnarray}

The third equality holds because of the Lagrangian
multiplier update rule:

$$W-{B^{(k+1)}}^*{L^{(k+1)}}^* =
\frac{1}{\beta^{(k)}}\left(
{\pi^{(k+1)}}^*-{\pi^{(k)}}^*\right)$$

Since ${{\pi^{(k)}}^*}$ is always bounded, we conclude that
\begin{equation}
\frac{1}{2}\mbox{tr}\left({B^{(k+1)}}^TB^{(k+1)}\right) -
\frac{1}{2}\mbox{tr}\left({B^*}^TB^*\right) \leq \mathcal{O}\left(\frac{1}{\beta^{(k)}}\right) \nonumber
\end{equation}
This completes the proof of the theorem.
\end{proof}

\section{Implementation of the Matrix Mechanism}\label{sec:appedinex:matrix}
In \cite{LHR+10}, Li et al. propose the \emph{Matrix Mechanism}. The core of their method is finding a matrix $A$ to minimize the following the program.
\begin{equation}\label{l2_app}
\min_{A\in R^{r\times n}} \|A\|_{2,\infty}^2\mbox{tr}(W^TWA^\dag
A^{\dag T})
\end{equation}
Li et al. \cite{LHR+10} present a complicated implementation that
may not be practical due to its high complexity. We hereby present a
simpler and more efficient solution to their optimization program.
Here $\|A\|_{2,\infty}^2$ denotes the maximum $\mathcal{L}_2$ norm
of column vectors of $A$, therefore
$\|A\|_{2,\infty}^2=\max(\mbox{diag}(A^TA))$. Since
$(A^TA)^{-1}=(A^TA)^{\dag}$ ($A$ has full column rank), we let
$M=A^TA$, and reformulate Formula (\ref{l2_app}) as the following
semidefinite programming problem:

\begin{equation}
\min_{M \in R^{n\times n}} \max(\mbox{diag}(M))  \mbox{tr}(W^TW
M^{-1}) ~  s.t.~ M \succ 0 \nonumber
\end{equation}

$A$ is given by $A=\sum_{i}^n \sqrt{\lambda_i} v_i v_i^T$, where
$\lambda_i, v_i$ are the $i$th eigenvalue and eigenvector of $M$,
respectively. Calculating the second term $\mbox{tr}(W^TW M^{-1})$
is relatively straightforward. Since it is smooth, its gradient can
be computed as $- M^{-1}W^TWM^{-1}$. However, calculating the first
term $\max(\mbox{diag}(M))$ is harder since it is non-smooth.
Fortunately, inspired by \cite{dAspremont2007}, we can still use a
logarithmic and exponential function to approximate this term.

\textbf{Approximate the maximum positive number:} Since $M$ is
positive definite, $v=\mbox{diag}(M)>0$. we let $\mu>0$ and define:
\begin{equation}\label{MatMac:Fobj}
f_{\mu}(v)=\mu \log
\sum_i^n\left(\exp\left(\frac{v_i}{\mu}\right)\right)
\end{equation}
We then have ${\max}(v) \leq f_{\mu}\left(v\right) \leq  {\max}(v) +
\mu \log n$. If we set $\mu=\frac{\epsilon}{\log n}$, this becomes a
uniform $\epsilon$-approximation of ${\max}(v)$ with a Lipschitz
continuous gradient with constant $\omega=\frac{1}{\mu} =
\frac{\log n}{\epsilon}$. The gradient of the objective function
with respect to $v$ can be computed as:
\begin{equation}\label{MatMac:Grad}
\frac{\partial f}{\partial v_i} =
\frac{\exp\left(\frac{v_i-{\max}(v)}{\mu}\right)}{\sum_j^n
\left(\exp\left(\frac{v_j-{\max}(v)}{\mu}\right) \right)}
\end{equation}

To mitigate the problems with large numbers, using the property of the
logarithmic and exponential functions, we can rewrite Eq.(\ref{MatMac:Fobj}) and Eq. (\ref{MatMac:Grad}) as:
$$f_{\mu}(v) = {\max(v)} + \mu \log \left( \sum_i^n
\exp\left(\frac{v_i-\max(v)}{\mu}\right)\right)$$
$$\frac{\partial f}{\partial v_i} = {\left(\sum_j^n
\exp\left(\frac{v_j-v_i}{\mu}\right)\right)}^{-1}$$

This formulation allows us to run the non-monotone projected gradient
descent algorithm \cite{Birgin2000} and
iteratively improves the result.

\end{document}